\documentclass[aps,pra,preprint,superscriptaddress]{revtex4-1}

\usepackage{graphicx}
\usepackage{amssymb}
\usepackage{amsmath}
\usepackage{bm}
\usepackage{verbatim}
\usepackage{color}

\newcommand{\derx}{\partial_x}
\newcommand{\der}{\partial}
\newcommand{\dd}{\mathrm{d}}
\newcommand{\ii}{\mathrm{i}}

\newcommand{\Rev }[1]{{\color{black}{#1}\normalcolor}} 
\newcommand{\Com}[1]{{\textbf{\color{red}{#1}\normalcolor}}} 

\begin{document}

\title[]{Quantum impurities: from  mobile Josephson junctions to depletons}

\author{Michael Schecter}
\affiliation{Center for Quantum Devices and Niels Bohr International Academy, Niels Bohr Institute, University of Copenhagen, 2100 Copenhagen, Denmark}

\author{Dimitri M. Gangardt}

\affiliation{School of Physics and Astronomy, University of Birmingham, B15 2TT, United Kingdom}

\author{Alex Kamenev}

\affiliation{School of Physics and Astronomy, University of Minnesota, Minneapolis, Minnesota 55455, USA}

\affiliation{William I. Fine Theoretical Physics Institute, University of Minnesota, Minneapolis, Minnesota 55455, USA}

\begin{abstract}
 We overview the main features  of mobile impurities moving in
 one-dimensional superfluid backgrounds by modeling it as a mobile Josephson
 junction, which leads naturally to the periodic dispersion of the impurity. 
 The dissipation processes, such as radiative friction and quantum 
 viscosity, are shown to result from the interaction of 
 the collective phase difference with the background phonons. We develop a more realistic depleton
 model of an impurity-hole bound state that
provides a number of exact results interpolating between the semiclassical weakly-interacting picture and  
the strongly interacting Tonks-Girardeau regime. We also discuss the physics of a 
trapped impurity, relevant to current experiments with ultra cold atoms.  
  
\end{abstract}

\maketitle

\section{Introduction}
\label{sec:intro}

The motion of mobile impurities in superfluid environments is a fascinating
subject with a long history. The field first came to prominence in the late forties
with experiments on $^4$He -- $^3$He mixtures. It was noticed that the super
flow through the supra-surface film does not involve He$^3$, leading to a
substantial purification of $^4$He leaking out of the container
\cite{PhysRev.72.502}. The phenomenon was initially attributed to the
absence of superfluidity in $^3$He. Soon after, Landau and Pomeranchuk
\cite{LandauPomeranchuk_1948}  realized that the effect has actually nothing to do
with the quantum statistics of the impurities, but rather with the fact that
foreign atoms cannot exchange energy and momentum with the superfluid
fraction. Instead, the rare impurities ought to contribute to the normal fluid
fraction. The nature of their interactions with the normal fraction was not
elucidated in the initial 1948 short paper \cite{LandauPomeranchuk_1948}, and was dealt with
in subsequent publications of Landau and Khalatnikov
\cite{LandauKhalatnikov1949ViscosityI,LandauKhalatnikov1949ViscosityII} and
Khalatnikov and Zharkov \cite{Khalatnikov_Zharkov_1957}. The latter authors realized that at
small temperatures the dominant interaction process is two phonon scattering
by $^3$He atoms, leading to impurity diffusion and equilibration with the
normal fraction. Since the scattering mechanism relies on the absorption of
thermal phonons, the diffusion coefficient is sharply divergent at small
temperature, $T$, and the corresponding linear in velocity, $V$, viscous
friction force scales as $F_\mathrm{fr}\sim T^8 V$. The theory was further
developed in a number of influential papers
\cite{,Bardeen_etal_PhysRevLett.17.372,Baym_PhysRevLett.17.952,Baym_PhysRevLett.18.71,Bardeen_etal_PhysRev.156.207,BaymEbner1967Phonon}
and verified experimentally through precision measurements of the velocity
and attenuation of sound \cite{PhysRevLett.17.1254}.  The subject was revived in the seventies in the
context of the storage of cold neutrons in superfluid $^4$He
\cite{Golub75,Golub77,Golub79}.

Recently the field has received growing attention due to advances in cold atom
experiments. Through a number of techniques it became possible to place
various impurity atoms in Bose-Einstein condensates (BEC) of alkali atoms,
manipulate their mutual scattering strength and apply forces selectively to
the impurity atoms. The Cambridge group \cite{Koehl_PhysRevLett.103.150601}
has used microwave pulses to flip the hyperfine state of a few spatially
localized atoms in the BEC of magnetically levitated $^{87}$Rb, turning them
into mobile impurities. The impurities, created in the hyperfine $m_F=0$
state, were then accelerated through the BEC by the gravitational force, not
compensated by the magnetic trap.  The Innsbruck \cite{PhysRevA.79.042718} and
Bonn \cite{Spethmann2012} groups have placed $^{133}$Cs impurities in a BEC of
$^{87}$Rb, and magnetically tuned their mutual scattering length with a
Feshbach resonance. The Florence group \cite{Catani2012} created mixtures of
$^{41}$K and $^{87}$Rb, and manipulated the two components with
species-selective optical potentials. Another line of research
\cite{Zipkes10,PhysRevLett.105.133201,Schmid_etal_PhysRevLett.105.133202}
deals with inserting a single {\em ion} into a BEC of neutral atoms using a
linear Paul trap to control the ion and study the mutual ion-atom
interaction. Although at the moment the ion micromotion leads to a continuous
depletion of BEC atoms from the trap
\cite{Zipkes10,Schmid_etal_PhysRevLett.105.133202}, this setup offers a
potential benefit in terms of easy manipulations with the help of
electrostatic fields.

One of the great advantages of the modern  ultra cold atomic experiments 
is the control over their  dimensionality 
by placing atoms into one, two or three dimensional
optical lattices. In particular, it has become possible to study impurity dynamics
in a one-dimensional (1D) atomic background, 
where the transverse motion is fully quantized
and only the lowest transverse sub-band is occupied by the atoms
\cite{Koehl_PhysRevLett.103.150601,Catani2012,Fukuhara2013}.  
The peculiarity of the 1D setup is that every
impurity atom effectively ``cuts'' the host liquid, creating an effective
 tunneling
Josephson junction (JJ) between the two superfluids. Unlike a conventional JJ,
however, the impurity is mobile and is characterized by its coordinate and
momentum, in addition to the Josephson phase, $\Phi$, across it. As we explain
below, the Josephson physics (and in particular the periodic dependence of
energy on $\Phi$) leads to a {\em qualitative} change of the impurity
dispersion relation, which goes far beyond a simple mass renormalization usually considered in higher dimensions.  The actual
energy-momentum relation $E(P,n)$ of a mobile impurity in a 1D superfluid with
density $n$ is a {\em periodic} function of the total momentum $P$ with the
period $2\pi\hbar n$.  This periodicity is due to the fact that in a system of size $L$ with $nL$ particles, the momentum
$nL\times(2\pi/L)=2\pi n$ may be transferred to the 1D
Galilean invariant host liquid with the energy cost
$nL\times(2\pi/L)^2/(2m)$, negligible in the $L\to\infty$ limit (here and below we set $\hbar =1$ and $m$ is the
atomic mass of the host superfluid). Therefore, the groundstate of a large
system whose momentum is an integer multiple of $2\pi n$ corresponds to a
super-flowing host and an impurity at rest with respect to it. 
We note that these considerations are not applicable in dimensions larger than one.

By following the dispersion curve $E(P,n)$ adiabatically through the application of a small external
force $F$ to the impurity, one expects to see Bloch oscillations with
period $2\pi n/F$ \emph{in the absence of any periodic lattice}. 
The mechanism behind these oscillations, first predicted in
Ref.~\cite{Gangardt09}, was attributed to the emergence of an effective
crystalline order of the background atoms, robust against thermal fluctuations for
sufficiently low temperatures as well as phonon radiation for sufficiently 
small external forces.

Although the dynamics of mobile impurities in 1D atomic condensates has
attracted a lot of attention
\cite{Gangardt09,Gangardt2010Quantum,Schecter_Gangardt_Kamenev_2012,Schecter2012,Mathy2012,Knap2014,PhysRevA.89.041601,PhysRevE.90.032132,PhysRevA.91.040101,PhysRevE.92.016101,PhysRevE.92.016102,Castelnovo2015},
a systematic pedagogical exposition of the consequences of the above mentioned
periodic dispersion is still missing.  This paper serves to fulfill this
gap. Here we investigate the dynamics of mobile impurities in a 1D quantum
liquid, exploring similarities and differences with the Josephson physics. Our
particular focus is on the conditions where the Bloch oscillations may be
observed. To this end we consider the thermal friction (i.e. due to the normal
fraction) along with the acceleration induced phonon radiation losses. We also
put a special emphasis on the consequences of being close to exactly
integrable points in the parameter space of impurity mass and impurity-host
interaction strength. An amazing consequence of dealing with Galilean
invariant 1D systems is that a number of exact results are available even away
from such integrable points. We will show below that the dispersion relation
$E(P,n)$, a {\em static} quantity available numerically or analytically in a
number of limiting cases, determines many {\em dynamic} characteristics
exactly, including those going beyond the linear response theory.  Finally, we
apply our results obtained for translationally invariant systems to the trap
geometry with an external adiabatic potential.  We give a number of estimates
for systems whose parameters are taken from recent experiments
\cite{Koehl_PhysRevLett.103.150601,Catani2012} as well as their immediate
extensions.

The paper is organized as follows: in Section~\ref{sec:josephson} we
illustrate the main ideas behind the physics of quantum impurities with a
simple, yet a non-trivial, model of a moving Josephson Junction (mJJ). In
Section~\ref{sec:coupling} we introduce the coupling of an impurity to the
background modeled as an elastic medium (phonons). We discuss the
mechanism of energy and momentum losses induced by such coupling and
derive the expression for the mobility. The simple model is then generalized 
to describe the impurity dynamics in any interacting quantum liquid at the expense of
introducing an additional coupling to density fluctuations in
Section~\ref{sec:general}. We illustrate this formalism in Section
\ref{sec:gp}  by deriving 
our previous  results for the impurity dynamics in a weakly interacting
background.    We then consider the background consisting
of impenetrable bosons (the Tonks-Girardeau gas) where the impurity
becomes a heavy polaron. We derive the first main
result of the paper for the  mobility of heavy polarons in
Section~\ref{sec:tg}. The second main result corresponds to the description of
impurities in a harmonically trapped background and is presented in
Section~\ref{sec:harmonic}. We conclude and discuss open questions 
in Section~\ref{sec:concl}.

\section{Moving Josephson Junction model}
\label{sec:josephson}

\begin{figure}[t]
  \centering
  \includegraphics[width=.7\columnwidth]{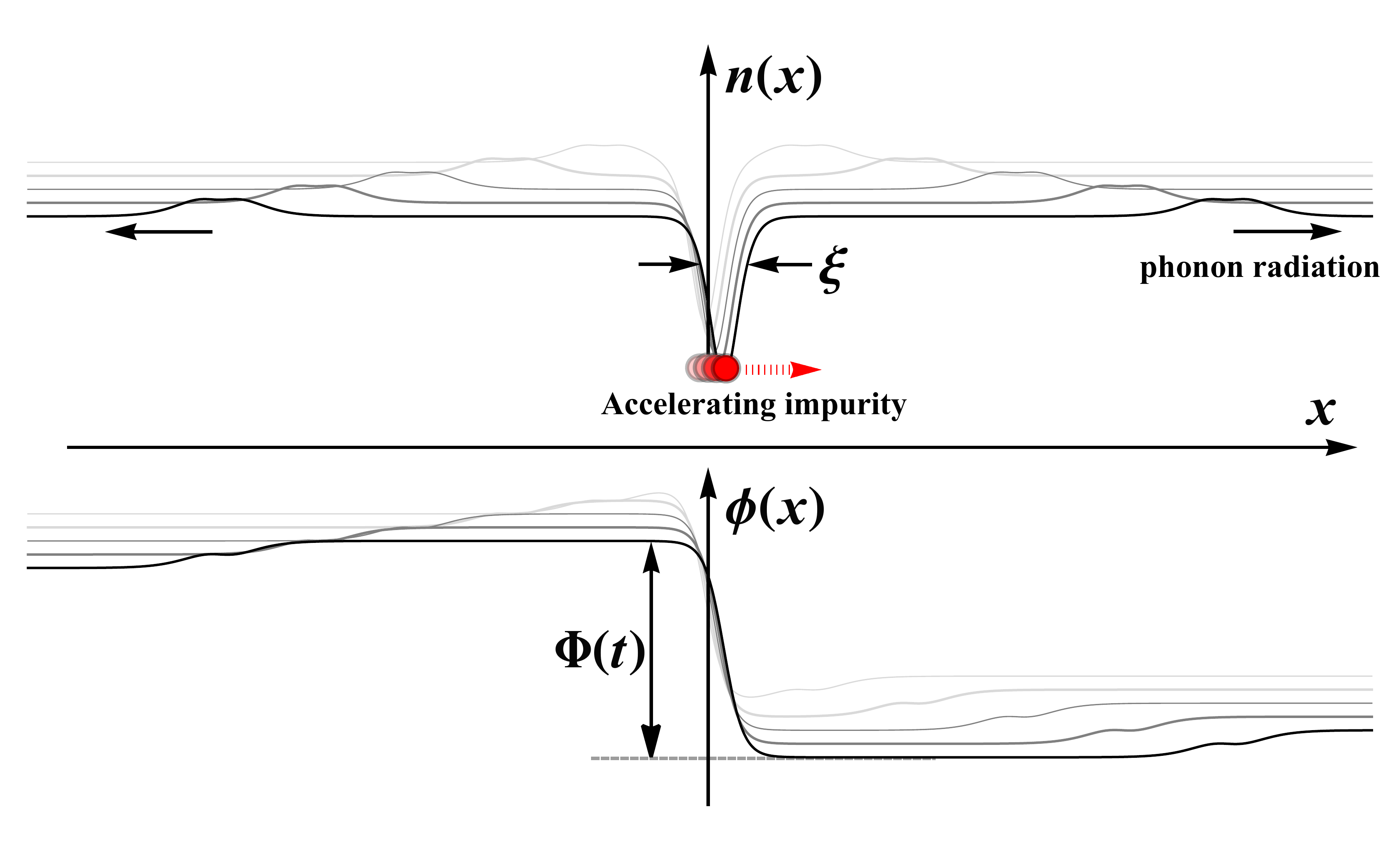}
  \caption{Mobile impurity modeled by a moving Josephson Junction. The interaction between the impurity and the host liquid creates a distortion of the local density and phase fields $n(x,t),\,\phi(x,t)$ of the host. When the impurity is driven out of equilibrium it excites phonons that propagate away at the sound velocity, and the phase drop $\Phi(t)$ becomes a dynamical quantity. The density and phase profiles are displaced vertically for clarity and represent, from top to bottom, snapshots of the fields as time evolves.}
  \label{fig:mJJ}
\end{figure}

The essential physics of a mobile impurity is most easily illustrated
by a strongly repulsive impurity moving in a background of weakly
interacting bosons. It can be modeled by a weak link located at the
position $X$ separating two  condensates
\footnote{According to the Bogoliubov-Mermin-Wagner theorem the true
  condensate is absent in one spatial dimension. Nevertheless, for our
  purposes the existence of a \emph{local} superfluid order is
  sufficient to define the phase difference across the impurity}.  
  The phase difference $\Phi$ between the two condensates, Fig.~\ref{fig:mJJ}, gives rise to the 
Josephson term in the energy 
\begin{eqnarray}
  \label{eq:JJ}
  H_\mathrm{d} (\Phi ) = - n {V}_\mathrm{c}\cos\Phi+\mu N
\end{eqnarray}
The  critical velocity is denoted by  ${V}_\mathrm{c}$, which depends on the impurity-background interaction. The last term in Eq.~(\ref{eq:JJ}) takes into account
the number of particles $N$ depleted by the impurity. Here we are working in
the grand-canonical ensemble with the chemical potential $\mu  \approx gn$ fixed by
the background density $n$ and interaction parameter $g$.

The phase drop $\Phi$ inevitably creates a small background
supercurrent $n\Phi/mL$. While the contribution
of the supercurrent to the total energy is of the order of $1/L$, its
contribution to the total momentum $P$ is independent of the system size, and is given by $n\Phi$. The total energy of the moving Josephson Junction 
(mJJ) is thus the combination of the Josephson term, Eq.~(\ref{eq:JJ}) and the
kinetic energy of the localized impurity,
\begin{eqnarray}
  \label{eq:JJplusB}
  H (P,X,\Phi) = \frac{1}{2\mathcal{M}} (P-n\Phi)^2 + U(X) + H_\mathrm{d}(\Phi). 
\end{eqnarray}
where $\mathcal{M} = M-mN$ is the total mass of the
impurity, including the  mass of $N$  particles it depletes from 
its vicinity. We have also included an external potential $U(X)$, e.g. of gravitational or optical origin, acting on the 
impurity.

The phase drop $\Phi$ represents a collective coordinate
characterizing the state of the impurity's depletion cloud.  In equilibrium its
value is determined from the requirement of 
the minimum of the total energy
(\ref{eq:JJplusB}): 
\begin{eqnarray}
  \label{eq:vel}
  (P-n\Phi)/\mathcal{M}=V_c \sin\Phi\, , 
\end{eqnarray}
The physical meaning of this condition is the matching between the  current
$I=nV$ of  the background particles  moving with velocity
$V=(P-n\Phi)/\mathcal{M}$ across mJJ  
and the Josephson current $nV_c\sin\Phi$.    Equation (\ref{eq:vel}) admits a solution $\Phi(P,n)$ that may be substituted into the Hamiltonian~(\ref{eq:JJplusB})
to obtain the dispersion curve $H(P,\Phi(P,n))=E(P,n)$ of the mJJ in the absence of the external potential $U$.
 Using the minimum condition 
Eq.~(\ref{eq:vel}) one may show that 
the velocity of the impurity satisfies
\begin{eqnarray}
  \label{eq:vel1}
V=(P-n\Phi)/\mathcal{M} = \der E/\der P \,, 
\end{eqnarray}
which defines the group velocity of the impurity dressed by the depletion cloud.
\begin{figure}[t!] \centering
   \includegraphics[width=1.0\columnwidth]{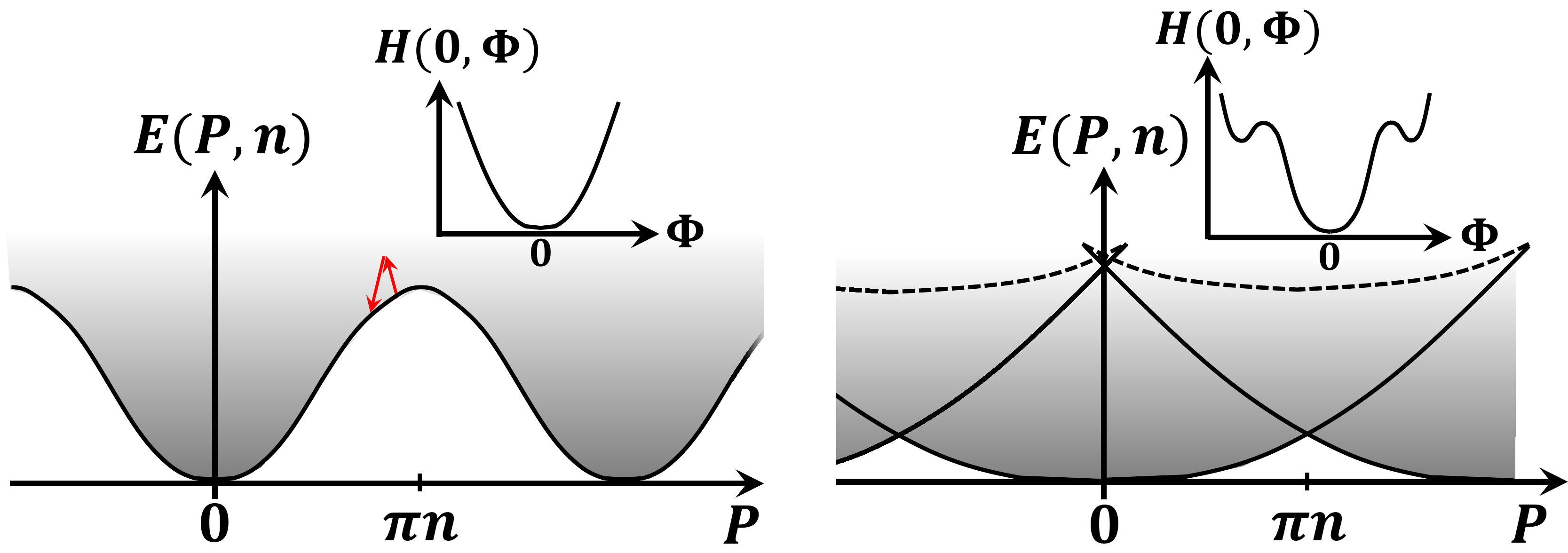}
   \caption{Schematic dispersion relation for a mobile  impurity in a 1D quantum liquid. Left panel: When the impurity mass is subcritical $M<M_c$ the dispersion is smooth, and the energy function $H(P,\Phi)$ has a unique minimum at $\Phi(P)$. At low temperatures, two-phonon scattering processes (red arrows) lead to energy and momentum relaxation of the impurity. Right panel: For $M>M_c$  the groundstate develops singular cusps at odd integer multiples of $\pi n$, and the function $H(P,\Phi)$ acquires metastable minima (the dashed lines in $E(P,n)$ represent local maxima of $H(P,\Phi)$).}
 \label{fig:dispersion}
 \end{figure}


One may notice a close similarity of the mobile impurity Hamiltonian (\ref{eq:JJplusB}) and the SQUID or phase qubit \cite{Devoret2004}. In this analogy $n^2/\mathcal{M}$ plays the role of the inductance of the SQUID loop, while the dimensional ratio $P/n$ is a direct analog of the external flux (in units of the flux quantum), permeating the loop. 
As in the case of the SQUID, the thermodynamic quantities  are periodic functions of the external flux with the period  $2\pi$, 
implying the {\em periodicity} of the dispersion relation   $E(P+2\pi n,n)= E(P,n)$, see Fig.~\ref{fig:dispersion}.
For example, in the case of a strong repulsive impurity, $V_c\ll c$, we have from  Eq.~(\ref{eq:vel}):
$\Phi(P,n) \approx P/n$ and $E(P,n) \approx  -nV_c \cos (P/n)+\mu N$.

Periodicity of the impurity's energy--momentum relation has
dramatic consequences for its dynamics: if the
momentum is linearly increased $P=Ft$  by an external force $F=-\partial_XU$, 
the velocity of the impurity does not increase indefinitely but 
changes \emph{periodically}, exhibiting Bloch
oscillations  with the period $\tau_\mathrm{B} = 2\pi n/F$, 
~\cite{Gangardt09, Schecter_Gangardt_Kamenev_2012}.  
This spectacular phenomenon is a close relative of the AC Josephson effect: 
under an applied constant force (voltage), the impurity velocity (current) is an oscillatory function of time. 
The mechanism is that once the time-dependent phase shift $\Phi(t)$ reaches $\pi$, the system undergoes a phase slip from $\pi\to-\pi$, which channels momentum $2\pi n$ into the superfluid background flow, and reverses the direction of the impurity's motion.

Another useful analogy is that of an impurity propagating in a periodic potential with the period $n^{-1}$ (this would be the case if the host gas forms a rigid  1D crystal). The energy spectrum of the impurity in such a lattice consists of Bloch bands periodic across the Brillouin zone with the width  $2\pi n$. 
Despite the fact that the background liquid is not actually a lattice, the {\em groundstate} energy of the liquid with an impurity is nevertheless a periodic function of the {\em total} momentum $P$, analogous to the lowest Bloch band in a periodic potential. The difference 
is that in the liquid there is a continuum of gapless excitations above the groundstate $E(P,n)$, which 
are due to the presence of the phononic modes. In the case of the rigid lattice, excited states at fixed momentum are separated by an energy gap, so the leading deviation from adiabaticity in the presence of an external force is given by exponentially weak Landau-Zener tunneling processes. The gapless modes of the superfluid background modify the adiabatic picture of Bloch oscillations in a much more substantial way. To capture the dynamics of a driven impurity we must generalize the static picture to the situation where $\Phi$ is a \emph{dynamical} variable. This is achieved in the next section by introducing the coupling of the impurity to phonons.

We mention that Eq.~(\ref{eq:vel}) may admit several distinct solutions  when the impurity mass exceeds a critical value (i.e. for $\mathcal{M}V_c/n > 1$). This corresponds to multiple metastable minima of the function $H(P,\Phi)$, which, for the case of a SQUID, represent trapped flux states in a system with large inductance. 
This feature has the distinguishing property that the groundstate is degenerate when the momentum is an odd multiple of $\pi n$ (the two states reflect the two independent solutions for $\Phi$ at this point). The corresponding level crossing leads to a cusp in the groundstate energy \cite{Lamacraft_PhysRevB.79.241105} as the momentum varies past $\pi n$, see Fig.~\ref{fig:dispersion}, and qualitatively changes the dynamics of a driven impurity,  as discussed in Ref.~\cite{Schecter2012}. 

Another remarkable phenomenon is the macroscopic quantum tunneling of phase 
between successive minima of  $H(P,\Phi)$ \cite{Caldeira1983Quantum}. It leads to the possibility of an impurity, trapped in such a meta-stable state, to transfer its energy and momentum to the host and thus experience an effective friction force $F_\mathrm{fr}$ even at zero temperature. Such a friction force appears to be a highly non-linear function of the impurity velocity  \cite{Astrakharchik2004Motion,Buechler_Geshkenbein_Blatter_2001}. It may seem to contradict the notion, discussed in the introduction, that only the normal fraction exerts friction on the impurity. The reason is that the condensate is, strictly speaking, absent in 1D even at $T=0$  due to long wavelength fluctuations of the phase. Moreover, once a heavy impurity reaches the lowest minimum of  $H(P,\Phi)$ it 
moves indefinitely (super flows) with a small velocity up to $\sim \pi n/M$, without any friction at $T=0$. A light impurity, 
$\mathcal{M}V_c/n <1$, does not exhibit metastable minima and is bound to relax to its only stable minimum $E(P,n)$, where it does {\em not} experience any $T=0$ friction, linear or non-linear.      

Another frequent misconception associated with a light mobile impurity, as
opposed to a static impurity or a tunneling barrier, is the
interaction-induced renormalization of its tunneling transparency. To make an
extreme version of the argument, consider an impurity in a repulsively
interacting Fermi gas. According to Kane and Fisher
\cite{Kane_Fisher_PhysRevB.46.15233} the tunneling transparency renormalizes
to zero in the limit of zero temperature, independent of the initial bare
value. This seemingly suggests that such an impenetrable impurity cannot move
and its dispersion must be flat.  The flaw in this argument is that the
Kane-Fisher renormalization is based on the $2k_F=2\pi n$ backscattering
processes, which for a finite mass impurity are associated with the recoil
energy $E_R=(2\pi n)^2/(2M)$. The renormalization thus terminates at this
finite energy scale \cite{castro96}, leaving the tunneling transparency and dispersion bandwidth finite. As a result, a
finite mass impurity has a non-flat ($2\pi n$ periodic) dispersion relation
$E(P,n)$ even in a repulsive Fermi gas.

Below, we focus on the experimentally most relevant case, where the mass is subcritical and the dispersion is a smooth periodic function of momentum, while  $H(P,\Phi)$ has a unique stable minimum at  $\Phi=\Phi(P,n)$. For the case of impurities with a supercritical mass we refer the reader to Ref.~\cite{Schecter2012}.

\section{Impurity-phonon coupling and dissipation}
\label{sec:coupling}

The static picture of the previous section needs to be modified if the
Josephson phase $\Phi$ becomes time-dependent. Since instanteneoous
changes of the phase in the left/right condensates are impossible,
one must take into account the generated gradients of the phase field,
\emph{i.e.} local currents which, in turn, lead to the density transport in
the form of phononic excitations, as illustrated in
Fig.~\ref{fig:mJJ}. 

For nonzero phononic fields,
the impurity is subject to the modified local supercurrent. The Josephson
Hamiltonian (\ref{eq:JJ}) should be modified by the tilting term   
\begin{eqnarray}
  \label{eq:phononint_0}
  H_\mathrm{int} = -\delta I \Phi\, ,
\end{eqnarray}
where $\delta I = (\delta\dot N_L - \delta\dot N_R)/2$ 
is the current through the
impurity, given by the rate of change of  the excess number of particles to
the left, $\delta N_L$, and to the right, $\delta N_R$, of the impurity.
Expressing these numbers via the integral of the density field,
\begin{eqnarray}
  \label{eq:ndens}
 \delta N_L = -\delta N_R =\int_{-\infty}^X \rho(x,t) dx 
= \frac{1}{\pi} \vartheta(X,t) ,
\end{eqnarray}
and using the standard bosonization definition \cite{GiamarchiBook} $\rho=\partial_x \vartheta/\pi$ of the field $\vartheta(x,t)$, we obtain
\begin{eqnarray}
  \label{eq:phononint}
  H_\mathrm{int} = -\frac{1}{\pi} \,\Phi \frac{\mathrm{d}}{\mathrm{d} t}\vartheta (X,t)= \frac{1}{\pi}\, \dot \Phi(t)  \vartheta (X,t)\, ,
\end{eqnarray}
where the full time derivative was omitted. Obviously, this term is only relevant for a time-dependent Josephson phase $\dot \Phi\neq 0$. Notice that it does not involve any coupling constants and thus represents  a universal coupling of the collective variable $\Phi$ to the phononic degrees of
freedom described by the field $\vartheta (x,t)$ and its canonical conjugate 
superfluid phase field $\varphi(x,t)$. Their dynamics can be
linearized  near equilibrium, resulting in the 
Luttinger liquid Hamiltonian \cite{GiamarchiBook}
\begin{eqnarray}
  \label{eq:LL}
  H_\mathrm{ph} =
   \frac{c}{2\pi}\int \!dx
 \left[ 
\frac{1}{K}
   (\derx \vartheta)^2
  +K \left(\derx\varphi\right)^2 \right]\,.
\end{eqnarray}
Here $K=\pi n/mc$ is the Luttinger parameter, proportional  to the compressibility of the background liquid.  For a weakly interacting superfluid $K\gg 1$, while for impenetrable bosons $K=1$.

We now integrate out the phononic degrees of freedom using the Keldysh technique \cite{KamenevBook} as
explained in Ref.~\cite{Schecter_Gangardt_Kamenev_2012}. As a result we obtain a quantum dissipative action,
similar to that of  the Caldeira-Leggett model \cite{Caldeira1983Quantum}. 
The dissipation arises naturally from the  continuous spectrum of phonons with a constant density of states at small energy, described by Eq.~(\ref{eq:LL}). This procedure results in a generically time non-local effective action for the impurity degrees of freedom $X(t)$ and $P(t)$, coupled to the collective variable $\Phi(t)$. 

\subsection{Zero-temperature dynamics and nonlinear mobility}
\label{sec:nonlinmob}

Postponing a discussion of fluctuation effects until Section
\ref{sec:fluctuations}, we focus here on the deterministic part of the
corresponding equations of motion, which are obtained by variation of the
effective action with respect to the ``quantum'' components \cite{Schecter_Gangardt_Kamenev_2012}
of $\Phi(t)$ and $X(t)$ degrees of freedom. The phase variable exhibits
over-damped dynamics with the effective ``friction'' coefficient $K/2\pi$,
\begin{align}
\label{overdampphase}
\frac{K}{2\pi}\, \dot\Phi &=-\frac{\partial H}{\partial \Phi}\, ,
\end{align} 
where the Hamiltonian $H$ is given by Eq.~(\ref{eq:JJplusB}). If the initial phase drop is off-equilibrium, it will evolve towards the value which minimizes $H$ by radiating away the excess phase difference in the form of phonons, see Fig.~\ref{fig:mJJ}. This results in an energy loss with the instantaneous rate
\begin{eqnarray}
  \label{eq:energloss}
 W=\frac{\der H}{\der \Phi} \dot\Phi  = -\frac{K}{2\pi}\, \dot\Phi^2 \, . 
\end{eqnarray}
In addition to the energy loss, the radiation of phonons also leads to the loss of momentum, i.e. 
a radiation friction force, $F_\mathrm{rad}$. The equation for the momentum for the mobile impurity becomes
\begin{align}
  \label{eq:friction_0}
  \dot P=F+ F_\mathrm{rad} &= -\frac{\partial U}{\partial X} - \frac{K}{2\pi} \frac{V}{c^2}\, \dot\Phi^2\, .
\end{align}
%
%
The main effects of the energy and momentum losses are to renormalize the period of oscillations $\tau_B$ and to
introduce a finite drift velocity $V_D$,
Ref.~\cite{Schecter_Gangardt_Kamenev_2012}. The latter can be
obtained from calculating the power radiated to the phononic bath averaged
over one period of oscillations and equating it to the work done by the
external force:
\begin{eqnarray}
  \label{eq:wbalance}
  FV_D =-\langle W\rangle_{\tau_B}= -\frac{1}{\tau_B}\int_0^{\tau_B} 
  \frac{K}{2\pi}\dot\Phi^2 \,  \dd t \, .
\end{eqnarray}
Using  $\dot\Phi = (\der \Phi/\der P)  \dot P$ and $\dot P\approx F$, 
we see that the drift velocity is proportional to the external force 
$V_D=\sigma F$. The proportionality coefficient is the {\em nonlinear} 
mobility $\sigma$, given by the integral over the Bloch oscillation period 
\begin{eqnarray}
  \label{eq:mob1}
  \sigma=\frac{1}{2\pi n}  
  \int_0^{2\pi n} 
  \frac{K}{2\pi}\left(\frac{\der\Phi}{\der P} \right)^2 \,  \dd P \, 
 \approx  \frac{K}{2\pi n^2}\, ,
\end{eqnarray}
where the last approximate equality is obtained assuming that  $\Phi\approx P/n$.
This result 
can be interpreted as an  inverse resistance using the 
analogy with  electrical current: in the
co-moving frame the impurity experiences current $I=nV_D$ and the
power dissipated on the impurity should be supplied by the
external force, $I^2 R = F V_D$, hence $1/R  =n^2\sigma = K /2\pi
$. It is exactly the electrical resistance  of a clean Luttinger liquid 
\cite{GiamarchiBook}, 
$R= h/e^2 K$, if one uses units such that $\hbar =e=1$ \footnote{ The inverse
proportionality to the Luttinger parameter is consistent with the fact
that the ``wire'' length is infinite and the effects of the ``leads''
discussed in Ref.~\cite{Maslov_Stone_PhysRevB.52.R5539} 
are irrelevant in our case}.    

\begin{figure}[t]
\centering
\includegraphics[width=0.6\columnwidth]{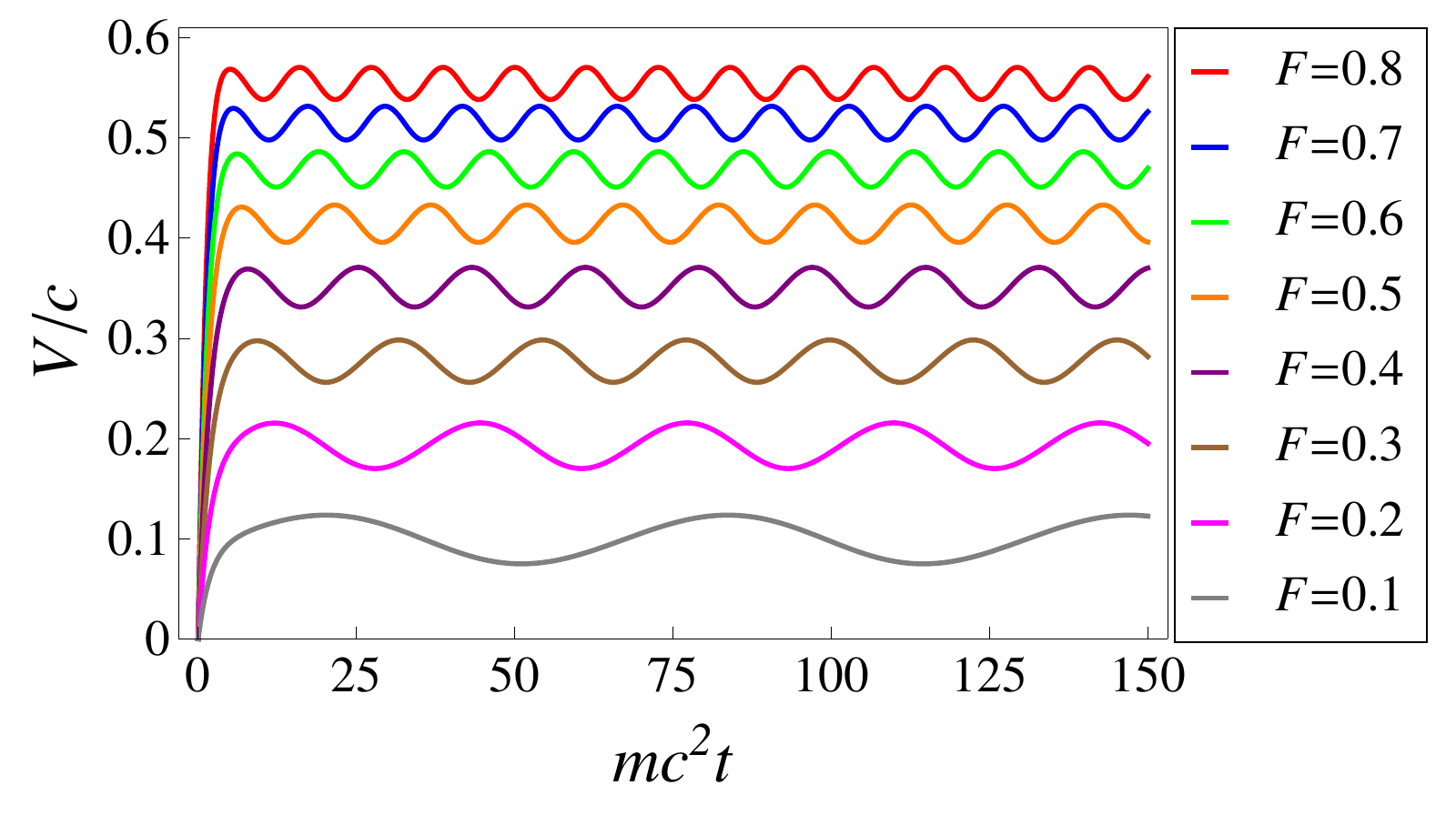}
\caption{Schematic velocity as a function of time for various forces listed in the
  legend ($F$ in units of $F_\mathrm{max}=2nmc^{2}$). As $F$ increases, the drift
  velocity and frequency of oscillations increases.}
\label{fig:V_vs_t}
\end{figure}

By no means should Eq.~(\ref{eq:mob1}) be interpreted in terms of
linear response theory: the drift motion of the impurity is
superimposed with the non-linear Bloch oscillations, see
Fig.~\ref{fig:V_vs_t}. 
The modified 
period of the oscillations can be calculated from the relation 
\begin{eqnarray}
  \label{eq:period}
  2\pi n = \int_0^{\tau_B} \dot P \, \dd t  = 
  \tau_B(F-\langle F_\mathrm{rad}\rangle_{\tau_B})\, . 
\end{eqnarray}
Using $\langle F_\mathrm{rad}\rangle_{\tau_B}/F  \approx  \sigma^2 F^2/c^2$
we obtain the renormalized period
\begin{eqnarray}
  \label{eq:taub}
  \tau_B = \frac{2\pi n}{F} \frac{1}{1-(F/F_\mathrm{max})^2}\, .
\end{eqnarray}
This expression can be trusted only for small forces $F\lesssim F_\mathrm{max}$, where the
characteristic force $F_\mathrm{max}$  is given by
\begin{eqnarray}
  \label{eq:critforce}
  F_\mathrm{max} = \frac{c}{\sigma} \approx 2mc^2n\, . 
\end{eqnarray}
Beyond this characteristic force the drift velocity exceeds the speed of sound $c$ and impurity emits Cherenkov radiation of phonons, which dramatically increases its energy and momentum losses. Since the Bloch oscillations do not take place in this regime, we shall not discuss it here.

We note that for the experiment of Ref.~\cite{Koehl_PhysRevLett.103.150601},
which used two hyperfine states of $^{87}$Rb for the impurities and background
gas, the coupling is rather strong $mg/n\sim 7$. In this case the mobility is
close to $\sigma=K/2\pi n^2\approx 1/2\pi n^2$. The external force is
provided by the gravitational field, which gives a drift velocity
$V_{\mathrm{D}}/c\sim 8$. The gravitational force thus exceeds the maximal
force $F_{\mathrm{max}}=2\pi^2 n^3/m$ by a factor of 8, and our low-energy theory is
inapplicable. The Bloch oscillations do not occur, and instead the impurities
become supersonic before exiting the gas. However, owing to the strong density
dependence of  $F_{\mathrm{max}} $, a gas twice as dense (or
sufficiently lighter, e.g. Li and Na) would provide a maximal force comparable
to the gravitational one and Bloch oscillations become possible, see
Table~\ref{tab:param}. The crossover between strong and weak force at
$F=F_{\mathrm{max}}$ was studied numerically in Ref.~\cite{Knap2014}, whose
results are consistent with our theoretical predictions.
\begin{table}[t!] 
\label{tab:param}
\centering
\begin{tabular}{c c c c c c} 
\hline\hline 
  & Li & Na & K & Rb & Cs \\ 
\hline 
$F_{\mathrm{max}}/F_{\mathrm{grav}}$\,\,\, & \bf{25} & \bf{1.7} & 0.54 & 0.12 & 0.05 \\ 
\hline 
$n_{\mathrm{crit}}\,[\mu\mathrm{m}^{-1}]$\,\,\, & 0.22 & 0.55 & 0.80 & 1.32 & 1.75 \\
\hline\hline 
\\
\end{tabular}\,\,\,\,\,\,\,\,\,\,\,\,\,\,\,\,\,\,\,\,\,\,\,
\caption{Typical parameters for various quantum gases in which impurities are created in the $m_F=0$ hyperfine state. Top row: Ratio of the maximal force over the gravitational force assuming a density of background particles $n=0.65\mathrm{\mu m}^{-1}$  in the strong coupling regime $\gamma\approx 7$ (as used in Ref.~\cite{Koehl_PhysRevLett.103.150601} for Rb) where $F_{\mathrm{max}}\approx 8 n^3/m$. At this density, Bloch oscillations are expected to occur for Li and Na, corresponding to $F_{\mathrm{max}}/F_{\mathrm{grav}}>1$ (bold entries). Bottom row: Critical density of various gases at fixed coupling $\gamma=7$. For $n>n_{\mathrm{crit}}$ Bloch oscillations are expected to occur.}
\end{table}

\subsection{Fluctuations}
\label{sec:fluctuations}

So far we have considered the zero temperature dynamics of an accelerated
impurity. We turn now to the finite temperature regime and focus on the
thermal fluctuations of the host liquid.  In doing so we shall assume that the
liquid is at thermal equilibrium with temperature $T$ in the laboratory
reference frame, and thus acts as a bath for the impurity.  In a generic
(non-integrable) case one expects that an excited impurity should thermalize
by losing its excess energy and momentum to the bath in the form of phonon
emission.

The problem, however,
is that due to the velocity mismatch, $V< c$, the emission of a single phonon is
energetically forbidden since $|E(P,n) - E(P\pm \omega/c,n)| < \omega$ (here $\pm$ refers absorption of a right/left moving phonon with energy $\omega$).  The leading process of energy and momentum exchange is therefore the {\em
  two-phonon} process. In this case the impurity first absorbs a thermal phonon
with energy $\omega\approx T $, bringing it to the virtual state
with momentum $P\pm \omega/c$, and then emits a Doppler shifted phonon with
the energy $\omega_{\pm}\approx \frac{c\mp V}{c\pm V} \omega$, see
Fig.~\ref{fig:dispersion}. One notices that, while both processes happen at the same
rate, there is a net momentum loss between them in the amount $(\omega_- -
\omega_+)/c \propto V\omega/(c^2-V^2)$. At small velocity $V\ll c$, this
implies a linear in velocity thermal friction force
\begin{equation}
\label{eq:friction}
F_{\mathrm{fr}}=-\kappa(T) V
\end{equation}
acting on the impurity. 

The above considerations indicate that: (i) since the two-phonon process relies on
thermal phonons, the friction coefficient $\kappa(T)$ is strongly temperature
dependent and vanishes at $T=0$; (ii) the thermalization process is not
uni-directional, but is rather diffusive with a drift in the momentum space.
Indeed, the same procedure of integrating out the phonons, described in Sec.~\ref{sec:coupling},
 leads naturally to the additional stochastic terms in the equations of
motion. They originate from the parts of the action that are quadratic in the
``quantum'' Keldysh components of the fields $\vartheta(X,t)$ and $\varphi(X,t)$, evaluated at the impurity coordinate. These fields can be conveniently decomposed into two independent
(chiral) auxiliary fields $\xi_\pm(t)$, whose equilibrium correlation functions
\begin{eqnarray}
  \label{eq:noise}
\Big\langle\xi_\pm(\omega)\xi_{\pm}(-\omega)\Big\rangle
=
K \omega\, \mathrm{coth}\frac{\omega}{2T_\pm}\, 
\end{eqnarray}
depend on the Doppler-shifted temperature $T_\pm =T(1\mp V/c)$. 
The corresponding equations of motions for the phase $\Phi(t)$ and momentum $P(t)$ are modified to become
\begin{eqnarray}
\label{eq:dotphinoise}
\frac{K}{2\pi}\, \dot\Phi&=&- \frac{\der H}{\der
  \Phi}+ \big(\xi_+ + \xi_-\big), 
\\
\dot P&=&F -\frac{K}{2\pi}\frac{V}{c^2}\,  \dot \Phi^2  +
\frac{1}{c}\,\dot \Phi\,  \big(  \xi_+ -  \xi_-\big).
\label{eq:fradnoise}
\end{eqnarray}
Notice that, since the impurity interacts with the liquid through the time
dependent phase shift, Eq.~(\ref{eq:phononint}), the stochastic term in its
equation of motion also comes with the multiplicative $\dot \Phi$ factor,
understood in the sense of Ito calculus.  The friction force, due to the
two-phonon processes discussed above, may be obtained from Eqs.~(\ref{eq:dotphinoise}), (\ref{eq:fradnoise}) as follows. We solve Eq.~(\ref{eq:dotphinoise}) as a frequency (time-derivative) expansion as
$ K\dot\Phi/2\pi =- \frac{\Gamma}{2}(\dot\xi_++\dot\xi_-)
  -\frac{\Gamma^2}{4}\left(\ddot\xi_++\ddot\xi_-\right)+\ldots\, , $  
  where $\Gamma^{-1} = -(\pi/K)\partial^2_\Phi H$, or according to
  Eq.~(\ref{eq:JJplusB}), $\Gamma\approx -K\mathcal{M}/\pi n^2$. 
Substituting
  this expansion into the last term of Eq.~(\ref{eq:fradnoise}) and averaging over the
  noise according to Eq.~(\ref{eq:noise}), one finds to the leading order in
  $V/c$
\begin{align}
  \label{eq:favg}
F_{\mathrm{fr}} =\frac{\pi \Gamma^2}{2Kc}
  \left[\langle \xi_+ \ddot\xi_+\rangle -\langle \xi_-\ddot\xi_-\rangle \right] 
  \simeq
  -\frac{\Gamma^2}{4c}
  \int_0^\infty\frac{\dd\omega}{2\pi}
  \frac{\omega^4}{\sinh^2\frac{\omega}{2 T} }\, \frac{V}{cT} 
  =-\frac{2\pi^3}{15 c^2}\Gamma^2 T^4 V. 
\end{align}
As a result the friction coefficient  in Eq.~(\ref{eq:friction}) is given by 
\begin{equation}
\label{eq:kappa}
\kappa(T) =  \frac{2\pi^3}{15 c^2}\,\Gamma^2 \,T^4.  
\end{equation}
The $T^4$ dependence of the friction coefficient in 1D, at low temperatures,
was first found by Castro-Neto and Fisher \cite{castro96}. This result is a
1D generalization of 3D Khalatnikov's $T^8$ result \cite{LandauKhalatnikov1949ViscosityI,LandauKhalatnikov1949ViscosityII,Khalatnikov_Zharkov_1957,Bardeen_etal_PhysRevLett.17.372,Baym_PhysRevLett.17.952,Baym_PhysRevLett.18.71,Bardeen_etal_PhysRev.156.207,BaymEbner1967Phonon}, mentioned in the Introduction.  We will show below
that, beyond the simple model discussed here, the amplitude $\Gamma$ may be
expressed {\em exactly} in terms of the impurity dispersion relation
$E(P,n)$. One can then check explicitly that for all known exactly solvable
models $\Gamma=0$, consistent with the idea that integrable systems do not
thermalize.
   
 At finite     
temperature we  therefore have two distinct regimes: for
$F<F_\mathrm{min} = \kappa(T) V_c$  Bloch oscillations do not occur and
after some initial acceleration the impurity attains a steady state
with the drift velocity $V_D = F/\kappa(T) =\sigma_\mathrm{Kubo} F$. In the low temperature regime considered here,  the linear Kubo mobility $\sigma_\mathrm{Kubo}\gg \sigma$ is large, see Table~\ref{tab:param2}.  In the range $F_\mathrm{min}<F<F_{\mathrm{max}}$, Bloch oscillations appear with the renormalized period
 $\tau_B = 2\pi n/\sqrt{F^2-F_\mathrm{min}^2}$, while the corresponding drift velocity is approximately given by  
$V_D\approx \sigma F + \sigma_\mathrm{Kubo} F_\mathrm{min}^2/2F$. As a result, the drift velocity is a non-monotonous function of the applied force with a sharp local maximum $V_D\approx V_c$ at $F\approx  F_\mathrm{min} $. Alternatively at a fixed force, the drift velocity is a non-monotonous function of temperature with a maximum attained when $ \kappa(T)=F/V_c$.

\begin{table}[t!] 
\label{tab:param2}
\centering
\begin{tabular}{c c c c c c}  
\hline\hline 
  & Li & Na & K & Rb & Cs \\ 
\hline 
$F_{\mathrm{min}}/F_{\mathrm{grav}}[\times 10^{-5}]$\,\,\,\, & 7.4  & 0.69 & 0.22 & 0.05 & 0.02 \\ 
\hline 
$\sigma[\mu\mathrm{m}^2/\hbar]$ & 2.4 & 2.4 & 2.4 & 2.4 & 2.4 \\
\hline 
$\sigma_{\mathrm{Kubo}}[\mu\mathrm{m}^2/\hbar\times10^5]$ & 1.5 & 1.5 & 1.5 & 1.5 & 1.5 \\
\hline\hline
\end{tabular}
\caption{Ratio of the minimal force over the gravitational force and mobilities for various quantum gases.  We assume the impurity is created in a distinct hyperfine state of the gas with the impurity-gas scattering length differing from the gas-gas scattering length by 10\% (when they are equal, or if the background gas is in the Tonks-Girardeau limit $F_{\mathrm{min}}=0,\,\sigma_{\mathrm{Kubo}}=\infty$ due to integrability). In all cases we have assumed a temperature $T=0.5mc^2$, density $n=0.65\mu\mathrm{m}^{-1}$ and coupling strength $\gamma=7$. Due to the closeness to integrability, the gravitational force always greatly exceeds the minimal force, and gives rise to Bloch oscillations if $F_{\mathrm{grav}}<F_{\mathrm{max}}$ (see Table~I).}
\end{table}

An additional consequence of the noise terms in
Eqs.~(\ref{eq:dotphinoise}),~(\ref{eq:fradnoise}) is  dephasing of
the oscillations even at zero temperature due to quantum fluctuations. Using the last term in Eq.~(\ref{eq:fradnoise}), together with $\Phi\approx P/n$, we have
\begin{eqnarray}
  \label{eq:phaserand}
  \Phi(t) \simeq 
  \frac{1}{n} \int_0^t dt' \dot P (t') \simeq \frac{F}{n}\left[ t + 
  \frac{1}{c}\int^t_0 dt' \big(\xi_+  -\xi_- \big) \right].
\end{eqnarray}
As a result, the oscillatory part of the noise-averaged velocity decays as a power law: 
\begin{eqnarray}
  \label{eq:velavg}
  \langle V(t) \rangle = V_D + V_c \langle \sin\Phi(t) \rangle  = V_D + 
  \frac{V_c}{(\mu t)^\alpha}  
  \sin \frac{Ft}{n}\, ,
\end{eqnarray}
where $\alpha = (4\pi/K) (F/F_\mathrm{max})^2$. 

\begin{figure}[!t]
\includegraphics[width=.45\columnwidth]{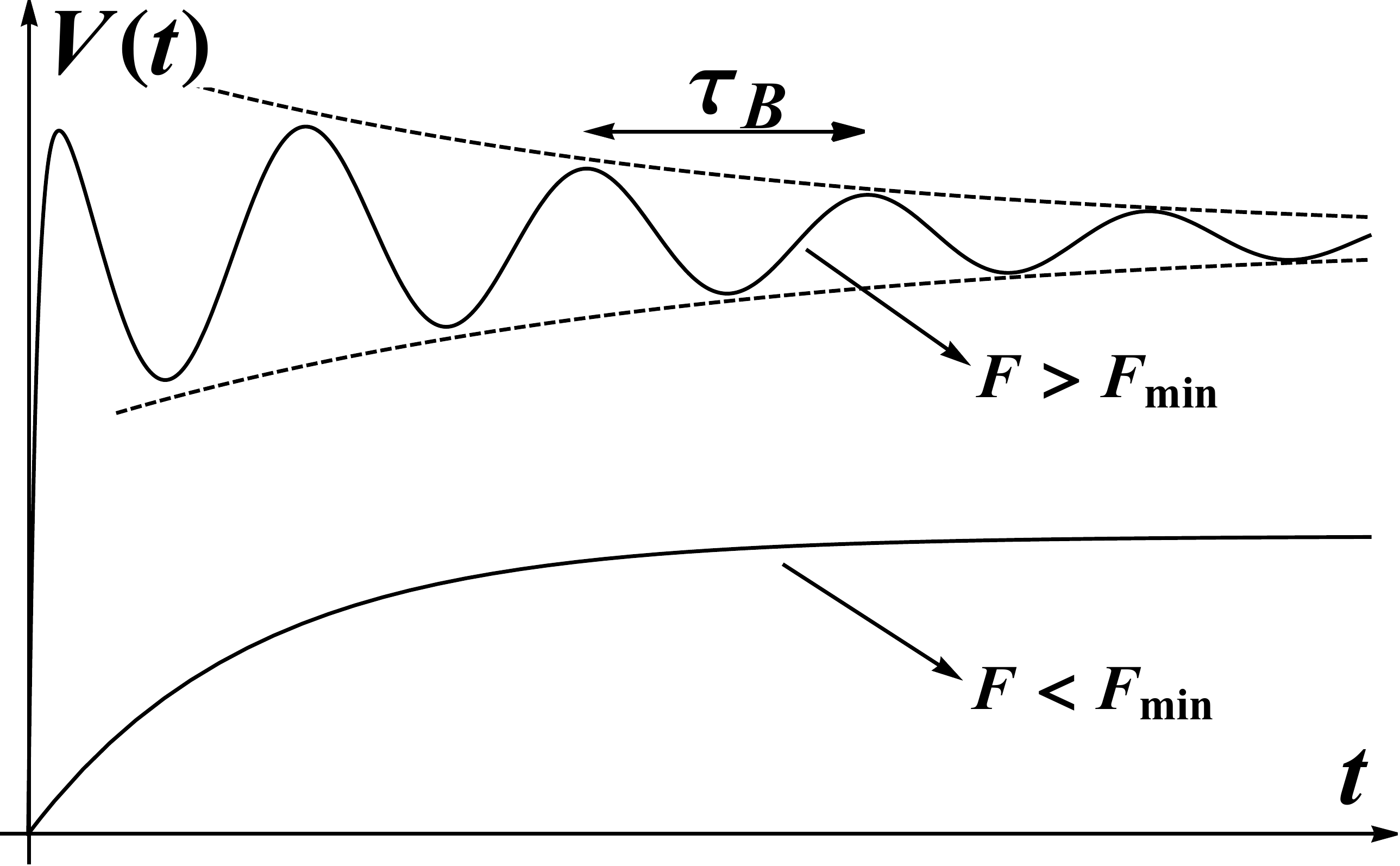}
\caption{Schematic noise-averaged velocity as a function of time including the effects of fluctuations. For $F<F_{\rm{min}}$ the impurity velocity saturates below the critical velocity and Bloch oscillations do not occur. For $F_{\rm{min}}<F<F_{\rm{max}}$ Bloch oscillations occur, but are attenuated in time due to dephasing, see Eq.~(\ref{eq:velavg}).}
\label{fig:velocity}
\end{figure}

The behavior of the
impurity velocity   is
illustrated in Fig.~\ref{fig:velocity}.  
At finite temperature the same calculation results in the exponential decay
for  the envelope of the Bloch oscillations:  $\langle V(t)\rangle\approx V_D
+V_c \exp (-\pi \alpha T t)\sin (Ft/n)$, which may lead to complete blurring
of Bloch oscillation phenomenon, in contrast to the power law dephasing at
$T=0$. 

\section{Mobile impurity  in a generic  superfluid background: the depleton model }
\label{sec:general}

 The phenomenology and the formalism, outlined above, are in no way
  restricted to the mJJ model. The generic description is obtained by
  acknowledging that in addition to the phase $\Phi(t)$ there is another
  collective degree of freedom, which may be chosen as the number of depleted
  particles $N(t)$.  The presence of two slow collective variables follows
  from the presence of two conservation laws: momentum and particle
  number. For a system conditioned to fixed values of $\Phi$ and $N$, all
  other degrees of freedom equilibrate quickly on the timescale $\mu^{-1}$ to
  form an optimal depletion cloud.  On the other hand, changing $\Phi$ and $N$
  is only possible by channeling momentum and particles into excitations of
  the liquid. When the time variation of $\Phi,\,N$ is slow (e.g. due to a
  small external force), these excitations consist of soft phonons whose
  wavelength greatly exceed the size of the depletion cloud $\xi$ \Rev{leading
    to the appearance of a fast time scale $\xi/c $. This time scale,
    being compared with the period of Bloch oscillations
    $\tau_B=2\pi n/F$, provides the upper bound on the external
    force $F$. This bound is identical to  the previously formulated condition 
    $F<F_\mathrm{max}$ and we use  such an adiabatic approach to 
    develop an analytically  tractable theory for the low-energy impurity dynamics
  \cite{Schecter_Gangardt_Kamenev_2012}. Recently this adiabatic approach was
  critisized in
  Refs.\cite{PhysRevA.89.041601,PhysRevE.90.032132,PhysRevE.92.016102}, based
  on the absence of the gap to lowest excitations. We note here, however, that
these excitations are phonons, traveling fast away from the depleton and thus
leaving it in a state of \emph{local} equilibrium, sufficient for using the
adiabatic approach.}

The Hamiltonian (\ref{eq:JJplusB}) is generalized to
\begin{eqnarray}
  \label{eq:HPhiN}
  H (P,X,\Phi,N) =\frac{1}{2} \frac{(P-n\Phi)^2}{M-mN}  +U(X)+\mu N+ H_\mathrm{d}(\Phi,N) \, .
\end{eqnarray}
The quantity $H_\mathrm{d} (\Phi,N)$ is the so-called depleton energy and is
constructed in such a way that the minimization of $H(P,\Phi,N)$ (without $U(X)$) with respect
to $\Phi, N$ for fixed momentum $P$ and density $n$ yields the equilibrium dispersion $E(P,n)$
of the impurity.
Conversely, if the exact groundstate energy $E(P,n)$ is known,  $\Phi(P,n)$ and $N(P,n)$ can be determined from the partial derivatives of $E(P,n)$ by solving the equations    
\begin{equation}
\label{eq:Phi N}
\frac{\der E}{\der P} = \frac{P-n\Phi}{M-mN}=V;\,\,\,\,\,\quad\quad
\frac{\partial E}{\partial n}=-V \Phi + \frac{mc^{2}}{n}N.
\end{equation}
The first of Eqs. (\ref{eq:Phi N}) is identical to Eq.~(\ref{eq:vel1}), while the second relation in Eq.~(\ref{eq:Phi N}) follows from taking the density partial derivative of Eq.~(\ref{eq:HPhiN}) in equilibrium, defined by $\partial_N H=\partial_\Phi H=0$. 

The equilibrium values of $N(P,n),\,\Phi(P,n)$ are also directly related to
the edge exponents of the impurity spectral function $A(P,\omega)$, which
represents the probability for an impurity with momentum $P$ and energy
$\omega$ to tunnel into the ground state of the liquid
\cite{Khodas07,Khodas2008PhotosolitonicEffect,Kamenev08,Imambekov08,Imambekov2009,Zvonarev_etal_PhysRevB.80.201102,Lamacraft_PhysRevB.79.241105,Imambekov2012}. Because
$E(P,n)$ defines the lower edge of the many-body spectrum in the presence of
the impurity, we have
$A(P,\omega)\propto\Theta(\omega-E(P,n))\left[\omega-E(P,n)\right]^{\beta(P,n)}$,
where $\beta=2K\left[(\Phi/2\pi)^2+(N/2K)^2\right]-1$. The power law behavior at the spectral threshold is a consequence of the orthogonality catastrophe and was discussed extensively in the review \cite{Imambekov2012} in terms of the phonon scattering phase shifts $\delta_\pm/\sqrt{\pi}=-\sqrt{K/\pi}\,\Phi\mp\sqrt{\pi/K}\,N$. These relations provide an interpretation of the phase drop $\Phi$ and the number of depleted particles $N$  beyond the semiclassical regime of weakly interacting bosons. Indeed, the phase shifts $\delta_\pm$ of the chiral low energy excitations  across a moving impurity may be defined for any interaction strength.

The coupling, Eq.~(\ref{eq:phononint}), must now be generalized to 
include the dynamics of  $N$. The form of the coupling remains universal  and is given by 
\begin{eqnarray}
  \label{eq:phononint_full}
  H_\mathrm{int} = \frac{1}{\pi}\,\dot \Phi \,  \vartheta (X,t) + \dot N  \varphi (X,t)\, .
\end{eqnarray}
Together with the Hamiltonian in Eq.~(\ref{eq:HPhiN}) the last equation
defines the depleton model.
Integrating out the phononic modes leads to the coupled dynamical equations for $P,X,\Phi,N$. Their solution in the limit $F\to 0$ yields the exact nonlinear mobility 
\begin{equation}
  \label{eq:mob}
  \sigma=\frac{1}{2\pi n}\int\limits_{-\pi n}^{\pi n}dP
  \left(\frac{c^2}{c^{2}-V^{2}}\right)
  \left[\frac{K}{2\pi}
    \left(\frac{\partial\Phi}{\partial P}\right)^{2}+
    \frac{V}{c}\left(\frac{\partial N}{\partial P}\right)
    \left(\frac{\partial\Phi}{\partial P}\right)
    +\frac{\pi}{2K}\left(\frac{\partial N}{\partial P}\right)^{2}\right].
\end{equation} 
For the thermal  friction force one finds  
\begin{eqnarray}
  \label{eq:friction2}
  F_\mathrm{fr}  =
-\frac{2\pi^3}{15c^2}\, \big| {\Gamma}\big|^2 
  \left(\frac{c^2+V^2}{c^2-V^2}\right) T^4 V\,.
\end{eqnarray}
It has the same form as Eq.~(\ref{eq:favg}) with 
the only difference that the  phononic backscattering amplitude $\Gamma$ depends on  derivatives of both
collective variables,
\begin{eqnarray}
\label{gamma12_Pn}
{\Gamma} (P,n) &=& -\frac{1}{c} \left(
\frac{M}{m}\frac{\der \Phi}{\der P}+  \Phi\frac{ \der N}{\der P}
- N\frac{\der \Phi}{\der P}
 +\frac{\der N}{\der n}\right).
\end{eqnarray}
As discussed below, the backscattering amplitude and therefore the thermal friction force  vanish for integrable models.

It is remarkable that finding  dynamical quantities, such as $\sigma$ and $\sigma_\mathrm{Kubo}$, only requires knowledge of 
 the dispersion $E(P,n)$, which is a purely thermodynamic quantity!  The latter may be evaluated 
in various limiting cases. The previously considered mJJ 
model can be obtained by considering a particle moving in 
a weakly interacting bosonic gas  by taking the limit of strong
repulsion between the impurity and the atoms in the background. The
latter can be modeled semiclassically by a Bose-Einstein condensate as we
explain in the next section. In this case the depleton parameters $N$ and 
$\Phi$ are obtained directly from the solution of the Gross-Pitaevskii equation.
In the case of a strongly interacting background, quantum fluctuations play a
dominant role and one has to use a full quantum-mechanical calculation 
for the dispersion $E(P,n)$. This can be done in the extreme
Tonks-Girardeau limit which is equivalent to free fermions as we show 
in  Section \ref{sec:tg}.

\section{Impurity in a weakly interacting  background}
\label{sec:gp}
For the case of a weakly interacting background 
the energy and momentum of the impurity
can be determined using the classical solution  $X(t)=Vt$ of the impurity's
coordinate. Here the the condensate wavefunction acquires  the traveling wave  form $\Psi(x,t)=\Psi(x-Vt)$ in the frame moving with the impurity and  satisfies the Gross-Pitaevskii equation (GPE)
\begin{eqnarray}
  \label{eq:GPE}
  -\ii V\derx \Psi = -\frac{1}{2m} \derx^2\Psi -
  g\left(n-|\Psi|^2\right)\Psi + G\delta(x)\Psi\, , 
\end{eqnarray}
where $g$ is the interaction coupling constant between the background atoms
and $G$ is the impurity-background interaction constant. Due to the presence of the repulsive contact interaction 
term, the moving impurity 
creates a depletion cloud which is  effectively bound to it.

\begin{figure}[t!]
\includegraphics[width=.49\columnwidth]{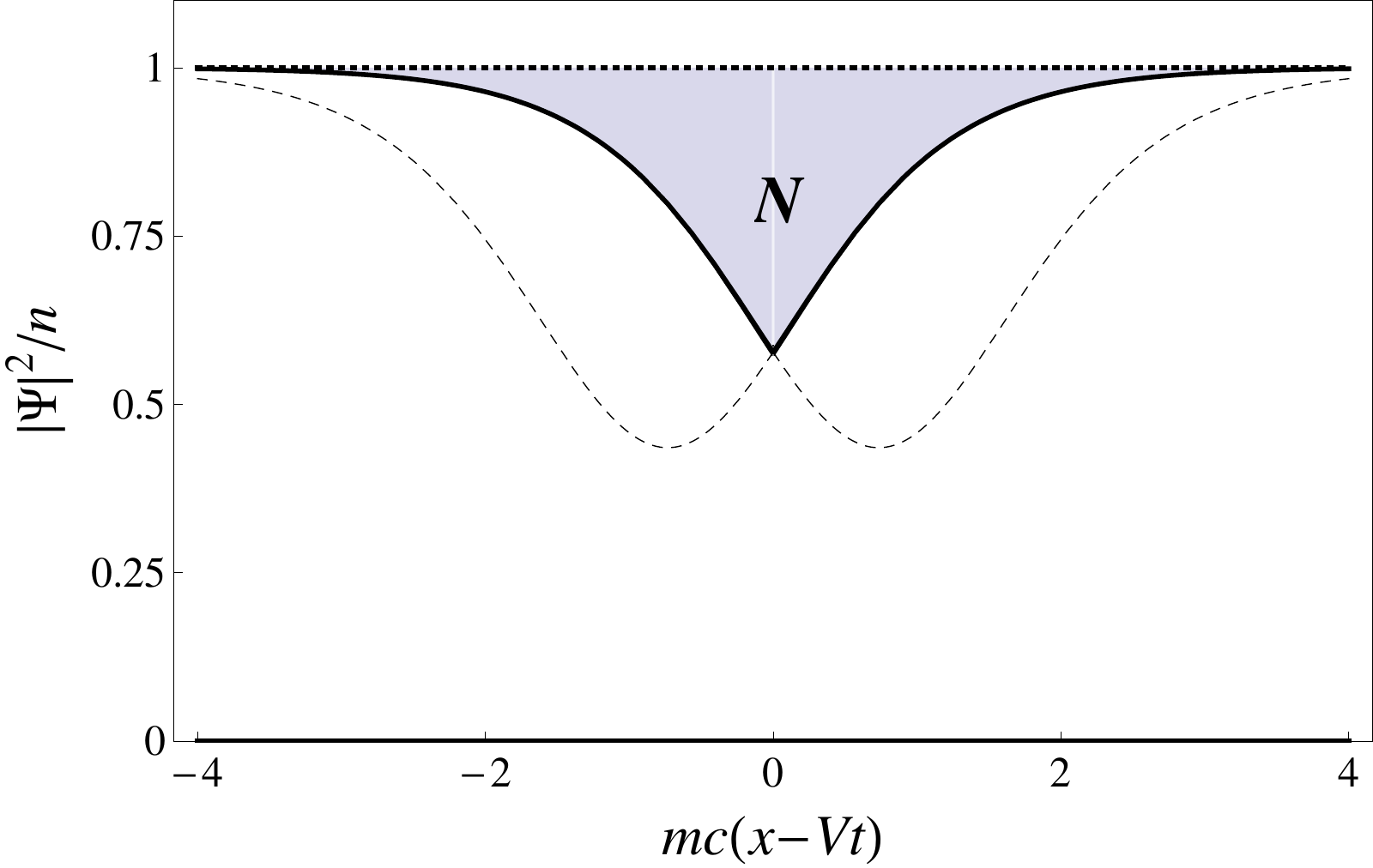}
\includegraphics[width=.49\columnwidth]{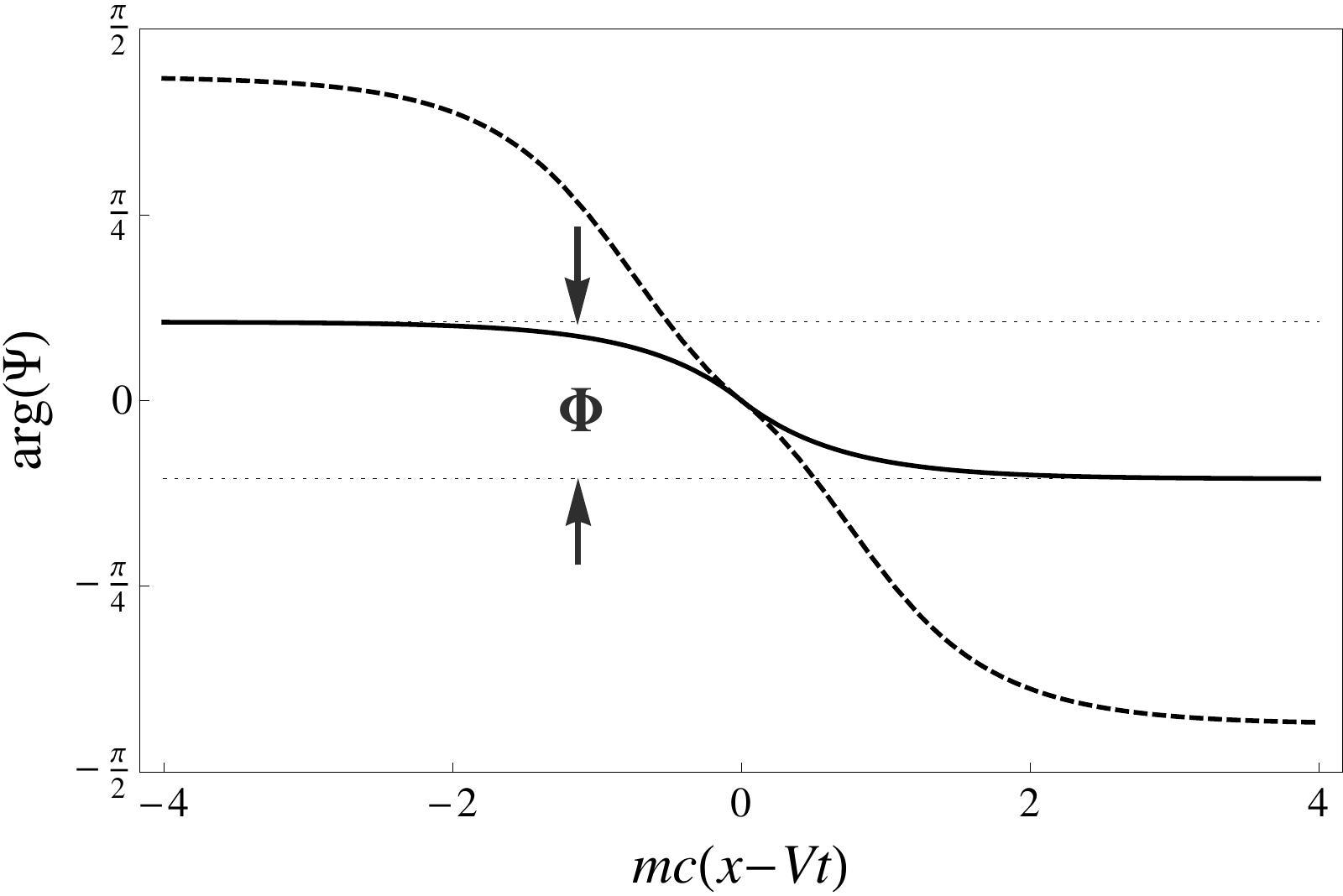}
\caption{Solution of Eq.~(\ref{eq:GPE}) obtained by matching two grey
  solitons. Left panel: the density profile. Right panel: the phase
  profile. The collective variables $N$ and $\Phi$ are shown.}
\label{fig:soliton}
\end{figure}

The shape of the depletion cloud can be obtained by constructing a solution from
two impurity free solutions (i.e. those with $G=0$) that satisfy the proper boundary conditions at location of the impurity:  $ \Psi'
(0^+)-\Psi'(0^-) = 2m G \Psi(0)$.  This strategy is facilitated by
the fact that for $V<c$ the bare GPE ($G=0$ in Eq.~(\ref{eq:GPE})) admits a one-parameter family of  soliton \cite{PitaevskiiStringariBook,Tsuzuki_1971} solutions:
\begin{equation}
\Psi_s(x)=
\sqrt{n}\left(\frac{V}{c}-\ii \sqrt{1-\frac{V^2}{c^2}}\, \textrm{tanh}
\frac{x}{l}\right),
\label{eq:wavefunction_no_impurity}
\end{equation}
where $l^{-1}= m\sqrt{c^2-V^2}$.  The solitons can
be visualized as a density dip having a core size $l$, as well as a corresponding phase drop. By appropriately matching two solitonic solutions at the impurity location one
solves Eq.~(\ref{eq:GPE}) \cite{Schecter_Gangardt_Kamenev_2012} as illustrated in Fig.~\ref{fig:soliton}.

From the solution $\Psi(x-Vt)$ the equilibrium values of the collective coordinates $N,\,\Phi$ can be computed directly in terms of the coupling $G$ and velocity $V$, as shown in Fig.~\ref{fig:soliton}. As expected, these values are in complete agreement with
 the thermodynamic definitions in Eqs. (\ref{eq:Phi N}). This can be shown by first solving for the energy $E(V,n)$ and momentum $P(V,n)$ as functions of $V$ using the equations
\begin{eqnarray}
\label{eq:energy1}
E&=& MV^2/2+
\int \dd x\left[ \frac{1}{2m}|\partial_x\Psi|^2+\frac{g}{2}\left(n-|\Psi|^2\right)^2\right]+G|\Psi(0)|^2,\\
\label{eq:mom1}	
P&=&MV+\ii\int \dd x
\Psi^{*}\partial_x\Psi +n\Phi\, .
\end{eqnarray} 
By inverting Eq.~(\ref{eq:mom1}) one finds $V(P,n)$, which can be substituted into the energy to yield the dispersion $E(V(P,n),n)=E(P,n)$. The same procedure independently yields the equilibrium values of the collective coordinates $N(P,n),\,\Phi(P,n)$ as functions of the total momentum, which allows one to check that the thermodynamic relations (\ref{eq:Phi N}) are indeed fulfilled.

With the impurity dispersion now in hand, one can proceed to compute the
nonlinear mobility $\sigma$ using Eq.~(\ref{eq:mob}) and the 
backscattering amplitude $\Gamma$ given by (\ref{gamma12_Pn}).
For a weak impurity, $G\ll c$, the main contribution to Eq.~(\ref{eq:mob})
comes from the regions of momentum where the velocity is maximal $V\approx V_c\approx c$,
leading to 
\begin{eqnarray}
  \label{eq:sigma_weak}
  \sigma \approx \frac{1}{nmG}\, , \qquad G/c \ll 1 \, .
\end{eqnarray}
which is enhanced compared to Eq.~(\ref{eq:mob1}) obtained for mJJ model. This
enhancement of mobility can be attributed to the fact that in the present case
the impurity is almost transparent to phononic excitations.

For calculation of the backscattering amplitude we can concentrate on $P\sim
0$ region and use the   perturbation theory in $G/c$  
to obtain $N\approx G/g,\,\Phi\approx P G/Mc^2$.  Then   Eq.~(\ref{gamma12_Pn})
leads  to the backscattering amplitude
\begin{eqnarray}
  \label{eq:gamma_weak_slow}
  \Gamma (P,n) = \frac{1}{mc^2} \left(\frac{G}{c}\right)\left(\frac{mG}{Mg}
    -1\right) \, ,
\end{eqnarray}
vanishing identically for the integrable case
$M=m$, $G=g$. 

In the case of a strongly repulsive impurity, $G\gg c$, the critical
  velocity $V_c=c^2/G\ll c $ is small and we have $N=2n/mc,\,\Phi=P/n$ for
  essentially any momentum $P$, due to the small bandwidth of the impurity
  dispersion. The nonlinear mobility 
\begin{eqnarray}
  \label{eq:sigma_weak}
  \sigma \approx \frac{K}{2\pi n^2}
  \left(1-\frac{1}{8}\frac{c^2}{G^2} \right)\, , \qquad G/c \gg 1 \, .
\end{eqnarray}
is only slightly different from the mJJ result Eq.~(\ref{eq:mob1}). 
 For the same reason the backscattering amplitude is approximately momentum
  independent and given by
\begin{eqnarray}
  \label{eq:gamma_strong_ds}
  \Gamma(P,n) = \frac{1}{mc^2} \left(1 - \frac{Mc}{n}\right)\, .
\end{eqnarray}
Assuming $Mc/n\ll 1$ leads to
$\Gamma=1/mc^2$ 
which coincides with the value $\Gamma=-K\mathcal{M} /\pi n^2$ derived
in Sec.~\ref{sec:fluctuations} for the mJJ model with 
$\mathcal{M} = - mN=-2n/c$.

\section{Impurity in Tonks-Girardeau gas}
\label{sec:tg}

In the case of a weakly interacting bosonic gas, the formation of the depleton
and its corresponding periodic dispersion law can be understood as a
consequence of the binding of a soliton to the impurity. The large number of
depleted particles $N\propto K\gg1$ allows one to develop a semiclassical
description of this binding, in which the density and phase fields can be
described using the mean-field Gross-Pitaevskii equation.

As the bosonic gas becomes more strongly interacting the number of missing
particles in the depletion cloud diminishes and the mean-field description
becomes inappropriate: both the soliton and its binding to the impurity must
be treated quantum mechanically. As long as the impurity mass is 
sub-critical  (see Section II) the transition from weak to strong coupling is a smooth crossover and the
impurity-soliton bound state remains intact. In this section we illustrate
this continuity by considering the extreme case of bosons with infinite
repulsion, widely known as the Tonks-Girardeau (TG) gas \cite{girardeau}.

One may represent the TG   gas of $nL$ hard-core bosons by free
fermions with momentum creation/annihilation operators satisfying $\{c_p,c^\dagger_{p^\prime}\}=\delta_{p p^\prime}$. This leads to the following Hamiltonian 
\begin{equation}
                                                                       \label{eq:H}
\hat H=-\frac{1}{2M}\,\frac{\partial^2}{\partial X^2} +\sum_p \frac{p^2}{2m} c^\dagger_p c_p + 
\frac{G}{L} \sum_{p,q} c^\dagger_pc_{p+q} e^{iqX}.
\end{equation}
We note that the above mapping to free fermions is valid for interactions of the density-density type, which we have assumed to be local in space.

To understand the low-energy properties of Eq.~(\ref{eq:H}), consider a state of the system with total momentum $P>0$. If
$P<P_0\equiv\mathrm{min}\{Mv_F,k_F\}$, the low energy states are those where
most of the momentum is carried by the impurity. Indeed, the impurity kinetic
energy $P^2/2M$ is less than that of soft particle-hole excitations above
the Fermi sea $\sim v_F P$. On the other hand, for $P>P_0$ the low energy
states are those where {\em hole excitations} carry a significant fraction of
the entire momentum $P$. The many-body ground state adiabatically connects
between these two limits, thus signaling strong impurity-hole hybridization at
$P\gtrsim P_0$.  As we show below, the strong hybridization manifests itself in the formation of an impurity-hole bound state. This non-perturbative process is responsible for the smoothness of the impurity dispersion relation, which in turn gives rise to Bloch oscillations under the application of an external force.

\begin{figure}
\includegraphics[width=.49\columnwidth]{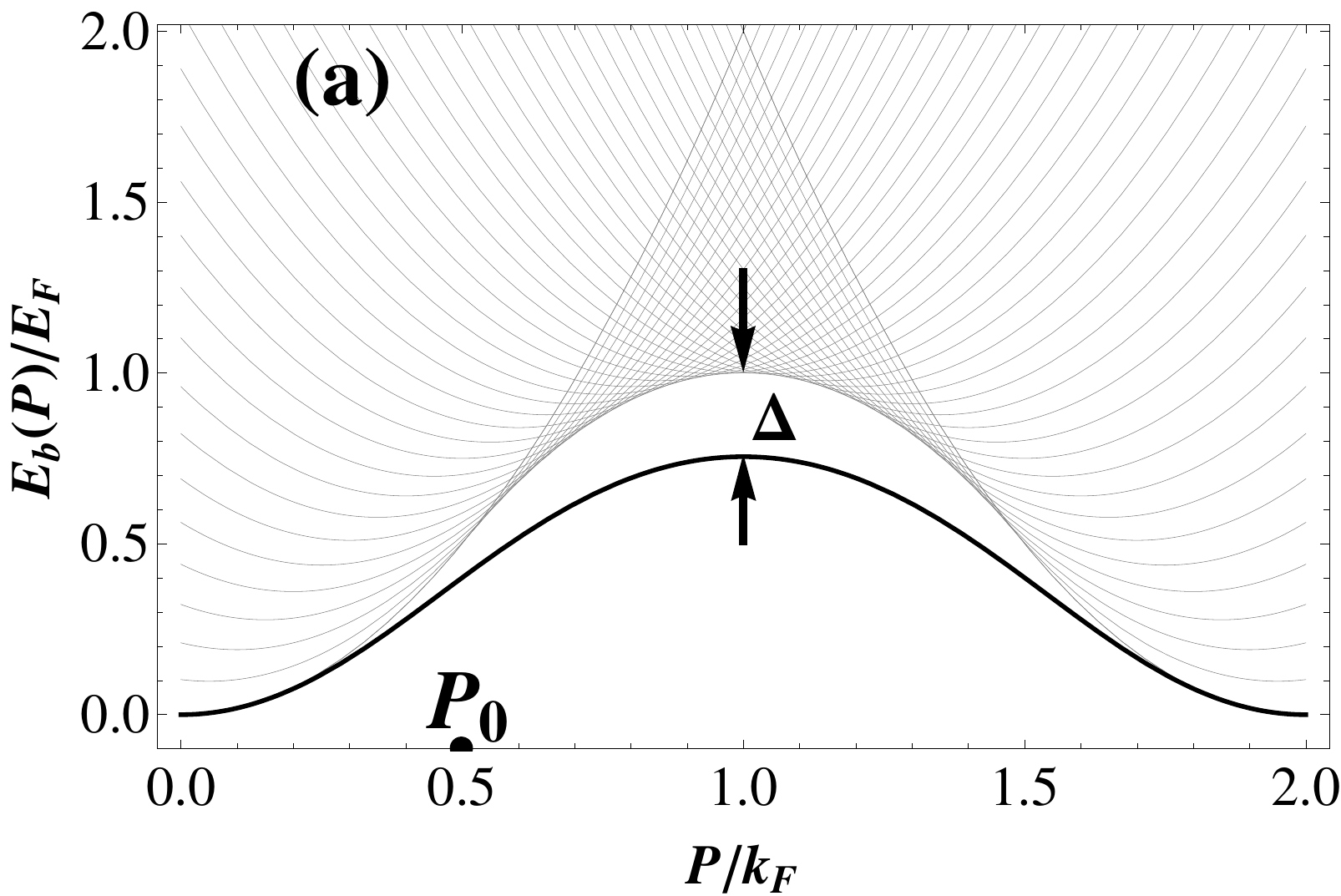}
\includegraphics[width=.49\columnwidth]{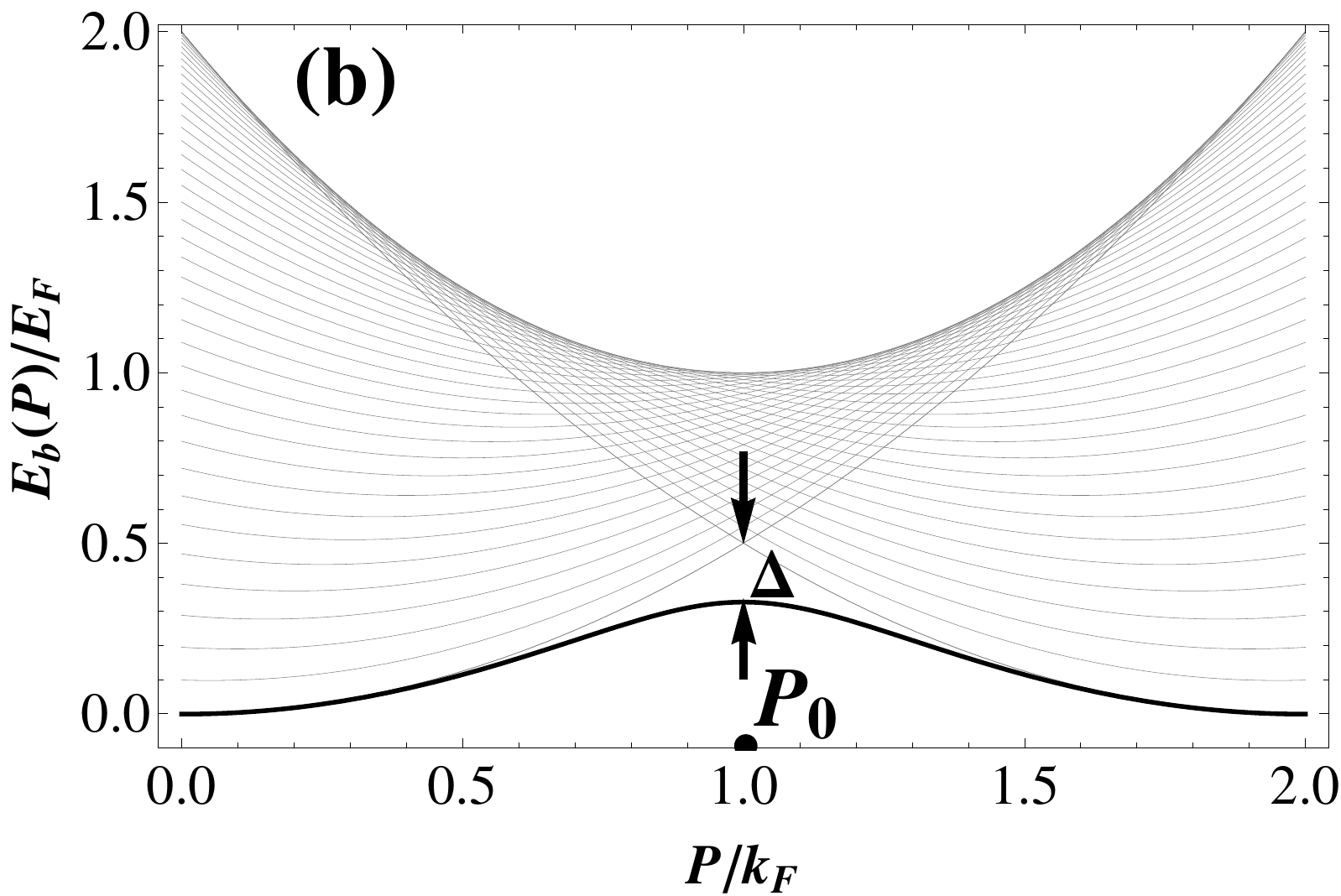}
\caption{(Color online) The bound-state $E_b(P)$, Eq.~(\ref{eq:integral}), (thick black line) and scattering continuum $\frac{(P- k)^2}{2M}+E_h(k)$ for a set of $k$ (thin gray lines) for the light impurity $M/m=1/2$ (a) and heavy impurity $M/m=2$ (b). In both cases $mG/n=0.7$.   }
\label{fig:fig1}
\end{figure}

To illustrate this effect, it is sufficient to 
consider a subspace of the full many-body space containing a single hole excitation with momentum $0<k<2k_F$, in addition to the
impurity with momentum $P-k$. This restriction is justified in the limit of weak coupling,
$G\ll v_F$, where the number of particle-hole pairs created by the impurity in the ground-state is suppressed. The basis vectors of this subspace are
\begin{equation}
                                                                     \label{eq:psi}
|k;P\rangle=  e^{i(P-k)X} c^\dagger_{k_F}c_{k_F- k}|\Psi_{\mathrm{FS}}\rangle\,,
\end{equation}
where $|\Psi_{\mathrm{FS}}\rangle$ denotes the unperturbed Fermi sea ground-state. The corresponding Schr\"odinger equation $\sum_{k'} \langle k;P| \hat H |k';P\rangle \psi_P(k') =E\psi_P(k)$ takes the form 
of a two-particle problem  with an {\em attractive} delta-interaction (formally the attraction arises from anti-commuting the fermionic operators in the last term in Eq.~(\ref{eq:H})),

 \begin{equation}
                                                                      \label{eq:Schrodinger}
\left[\frac{(P- k)^2}{2M}+\!E_h(k)+nG\right] \psi_P(k) - G\!\! \int\limits_{0}^{2k_F}\! \frac{dk'}{2\pi}\,\psi_P(k')\!=\!E\psi_P(k).  
\end{equation} 
Here $E_h(k)=v_Fk-k^2/2m$ is the hole kinetic energy (we measure $E$ relative to $NE_F/3$). This problem admits a unique bound-state solution, whose energy $E=E_b(P)+nG$ is found from the integral equation  
 \begin{equation}
                                                                      \label{eq:integral}
\int\limits_{0}^{2k_F} \frac{dk'}{\frac{(P- k')^2}{2M}+E_h(k') -E_b(P)} = \frac{2\pi}{G}\,. 
\end{equation} 
The resulting bound-state dispersion, shown in  Fig.~\ref{fig:fig1}, is a smooth periodic function of the total momentum, which is split from the scattering continuum $E_h(k)+ (P- k)^2/2M $ by the gap $\Delta$.

The hard gap between the bound-state and the continuum is an artifact of
restricting the particle in Eq.~(\ref{eq:psi}) to be created right at the
Fermi momentum $k_F$. Allowing for slight deviations $c^\dagger_{k_F} \to
c^\dagger_{k_F+p}$, enlarges the Hilbert space to include,
in addition to the bound-state,   low energy,
$\sim v_Fp$, particle-hole  excitations.  It is well
known  \cite{Kamenev08,Imambekov2012,Lamacraft_PhysRevB.79.241105}
that  interactions with these excitations 
transforms the bound-state into the {\em quasi}
bound-state with the power-law (instead of the pole) 
spectral function $A(P,\omega)$. 
These low energy excitations are also responsible for radiation
losses and thus for the finite mobility $\sigma$. 
As long as the external force is sufficiently small, $F<F_{\mathrm{max}}$, they do not destroy the Bloch oscillations associated with the impurity following the quasi bound-state. 

\subsection{Results for the exactly integrable model $M=m$}

It is worth noticing that the one-hole bound-state solution
(\ref{eq:integral}) is in {\em quantitative} agreement with the available
exact results.   For example, for $M=m$, the integrability of the model given
by  Eq.~(\ref{eq:H}) allows one to determine the exact  ground-state energy $E(P,n)$   \cite{Lamacraft_PhysRevB.79.241105}. It is defined implicitly through the integral relations
\begin{eqnarray}
\label{eq:BA-dispersion}
\nonumber
E(\Lambda)&=&\frac{k_F^2}{2m}+\int\limits_{-k_F}^{k_F} \frac{dk}{2\pi}\frac{4mG}{(mG)^2+4(k-\Lambda)^2}\left[\frac{k^2}{2m}-\frac{k_F^2}{2m}\right],
\\
P(\Lambda)&=&-2\int\limits_{-k_F}^{k_F} \frac{dk}{2\pi}\mathrm{arctan}\frac{2(k-\Lambda)}{mG},
\end{eqnarray}
where one must eliminate $\Lambda$ in the upper equation using $\Lambda(P)$ from the lower equation.

In the vicinity of $P\sim k_F$, where the one-hole bound-state is expected to be valid, one finds $E(P\to
k_F)=E_F-\frac{2\pi v_F}{3G}\frac{(P-k_F)^2}{2m}$. One may indeed verify from
Eq.~(\ref{eq:integral}) that $E_b(P\to k_F)+nG\approx E(P\to k_F)$. The effective mass of the bound-state, $M^*=-\frac{3mG}{2\pi v_F}$, therefore agrees with the exact result (up to perturbative corrections of $\mathcal{O}(G^2)$ which are subleading for $G\ll v_F$).  This shows that the single hole binding to the impurity is indeed 
the leading physical effect in the weak coupling limit. 

At strong coupling the impurity becomes dressed by multiple particle-hole
pairs and the above one-hole ansatz loses its quantitative
applicability. Nevertheless, the concept of the depleton as the 
impurity-hole bound state is
expected to survive, in the sense that the impurity drags with it a depletion
cloud with precisely one missing particle (i.e., a localized hole). This
statement can be made precise by studying the ground-state pair correlation
function $\langle n(x) n_i(0)\rangle$, which measures the fermion density a
distance $x$ away from the impurity (here $n_i$ is the impurity density
operator).

For the integrable case $M=m$ the pair correlation function was studied analytically by McGuire \cite{McGuire_1965}, with the strong coupling $G\gg v_F$ result 
\begin{eqnarray}
\label{eq:pair-corr}
\langle n(x) n_i(0)\rangle=n\left(1-\frac{\mathrm{sin}^2k_F x}{k_F^2 x^2}\right).
\end{eqnarray}
Integrating the deviation of Eq.~(\ref{eq:pair-corr}) from the background density $n$ over all space yields the number of depleted particles $N=\int_x \left(n-\langle n(x) n_i(0)\rangle\right)=1$. 

McGuire also studied the pair correlation function and ground-state energy $E$ for arbitrary coupling, in the case of zero momentum $P=0$. It is interesting to note that from McGuire's solution the number of depleted particles, as defined through the pair correlation function $\int_x \left(n-\langle n(x) n_i(0)\rangle\right)$, is identical to the thermodynamic expression $\partial_\mu E$ ($\mu=\frac{k_F^2}{2m}$ is the chemical potential of the background fermions)
\begin{eqnarray}
\label{eq:N-corr}
N=\partial_\mu E=\int_x \left(n-\langle n(x) n_i(0)\rangle\right) = \frac{2}{\pi}\mathrm{arctan}\frac{G}{2v_F}.
\end{eqnarray}
This result substantiates our intuition that $N$, as defined through the thermodynamic relation (\ref{eq:Phi N}) (at $P=0$ in the present case), is indeed related to the real space depletion of particles in the vicinity of the impurity, despite the absence of its semiclassical description. The corresponding  lengthscale $\xi$ of the depletion cloud is of course just the Fermi wavelength, cf. Eq~(\ref{eq:pair-corr}), in agreement with the general expectation $\xi=1/mc$  \cite{Schecter_Gangardt_Kamenev_2012} 

\subsection{Exact Nonlinear Mobility}

\begin{figure}
\includegraphics[width=.70\columnwidth]{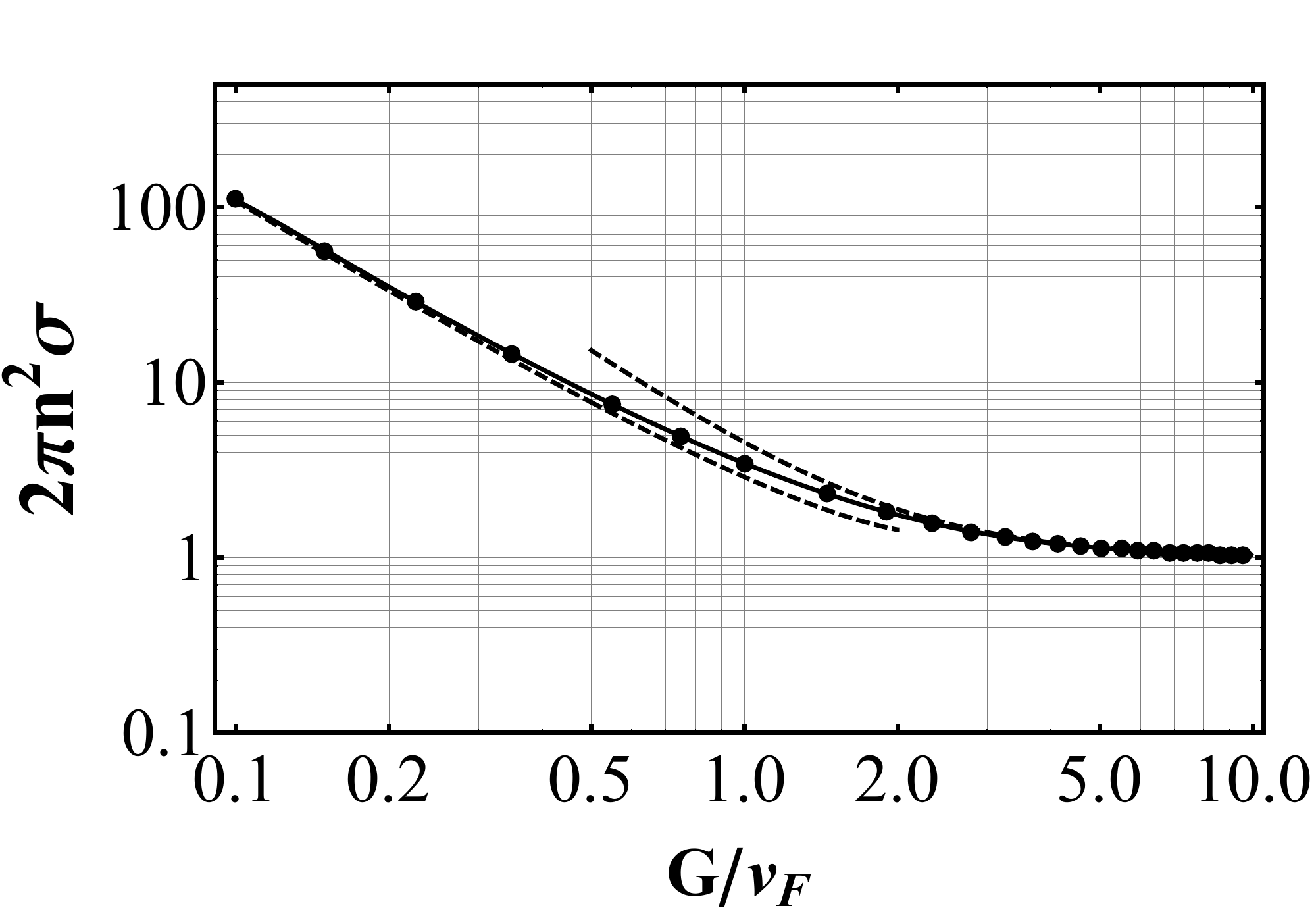}
\caption{Nonlinear mobility for the equal mass impurity in a Tonks-Girardeau gas. The dashed lines are the asymptotic limits given by Eq.~(\ref{eq:BA-mob}).}
\label{fig:mob}
\end{figure}

The above results confirm the idea that the ground-state properties of the model can be understood in terms of the impurity-hole bound state. The dynamic response of the bound-state can be described within the depleton framework of Sec.~\ref{sec:general}, where it was discussed that the response to an external force $F$, can be characterized by the purely thermodynamic quantity $E(P,n)$. By computing the dispersion from the integral equation (\ref{eq:BA-dispersion}) we may determine the exact mobility using Eqs.~(\ref{eq:Phi N}), (\ref{eq:mob}). At weak or strong coupling the mobility is given by
\begin{eqnarray}
\label{eq:BA-mob}
\sigma=\frac{1}{2\pi n^{2}}\begin{cases}
\frac{4v_{F}^{2}}{G^{2}\mathrm{ln}\frac{4v_{F}}{G}}, & G\ll v_{F};\\
1+\frac{32}{9}\frac{v_{F}^{2}}{G^{2}}, & G\gg v_{F}.
\end{cases}
\end{eqnarray}
These asymptotic formulae provide rather tight bounds on the exact mobility deduced numerically from the integral equations (\ref{eq:BA-dispersion}), as shown in Fig.~\ref{fig:mob}. One can arrive to Eq.~(\ref{eq:BA-mob}) in the limit of strong coupling $G\gg v_F$ by expanding the functions $V(P,n),\,N(P,n),\,\Phi(P,n)$  to the leading order in $v_F/G$.
Substitution of the resulting expressions into Eq.~(\ref{eq:mob}) gives the second line of Eq.~(\ref{eq:BA-mob}).

In the limit of weak coupling the dispersion acquires a more complicated form: it consists of essentially unperturbed parabolae centered at momenta $P=2jk_F$ for integer $j$ with  weak anti-crossings at $P=(2j+1)k_F$. The value of the collective coordinates $N(P,n),\,\Phi(P,n)$ thus remain close to zero at small momentum $P<k_F$ and change rapidly to $(N,\Phi/\pi)\to 1$ in the vicinity of $k_F$ in a window of width $mG$. The momentum derivatives $(\partial_P N,\partial_P \Phi/\pi)$ which enter the mobility formula (\ref{eq:mob}) are  strongly peaked at $P=k_F-mG/2$, with height $\propto 1/mG$. One may thus approximate for $G\ll v_F$
\begin{eqnarray}
\label{eq:mob-weak}
\nonumber\sigma&\approx& \int_{k_F-mG/2}^{k_F} \frac{dP}{n}\frac{(\partial_P N)^2}{1-V^2/v_F^2}\sim \frac{1}{nmG}\frac{1}{1-V^2(k_F-mG/2)/v_F^2}
\\
&\sim & \frac{1}{n^2}\frac{v_F^2}{G^2\mathrm{ln}(4v_F/G)},
\end{eqnarray}
where $V(k_F-mG/2)/v_F\approx 1-\frac{G}{2\pi^2 v_F}\mathrm{ln}\frac{4v_F}{G}$
can be obtained from second order perturbation theory, see e.g. Eq.~(2) of
Ref.~\cite{Lamacraft_PhysRevB.79.241105}. Keeping track of the numerical prefactor in
Eq.~(\ref{eq:mob-weak}) leads to the first line of
Eq.~(\ref{eq:BA-mob}). Deviations away from the integrable point $M=m$ do not
significantly affect Eq.~(\ref{eq:BA-mob}) provided $|1-M/m|<G/v_F$. As we
shall see in the next section, however, the backscattering amplitude, and thus
the Kubo mobility, is strongly sensitive to the deviation from integrability.

\subsection{Backscattering amplitude}
As shown previously \cite{Gangardt09,Schecter_Gangardt_Kamenev_2012} the backscattering amplitude vanishes at points of exact integrability. For the TG gas this implies that $\Gamma\propto 1-M/m$ when $M\sim m$. Below we verify this behavior and obtain the exact prefactor in various limiting cases where the analytic form is available.

In the limit of strong coupling we may set $N=1$ and neglect terms proportional to $1/M^*\propto 1/G$ in Eq.~(\ref{gamma12_Pn}). We then find
\begin{eqnarray}
\label{eq:Gamma-strong}
\Gamma=-\frac{\pi}{mv_F^2}\left(\frac{M}{m}-1\right),\,\,\,G\gg v_F.
\end{eqnarray}
This result is independent of momentum to leading order, owing to the essentially flat dispersion with bandwidth $\propto 1/G$.

At small coupling the backscattering amplitude acquires a complicated momentum dependence and we restrict ourselves to its behavior in the vicinity of the analytically accessible points $P=0,\,k_F$. At small momentum one may resort to second order perturbation theory to obtain $E(P,n)\approx P^2/2M^*+\mu_d$ with $M^*(P=0)=M\left(1+(G/\pi v_F)^2\right)$ and $\mu_d=nG$. Substituting this dispersion into Eq.~(\ref{gamma12_Pn}) gives
\begin{eqnarray}
\label{eq:Gamma-weak1}
\Gamma(P=0)=-\frac{2\pi}{mv_F^2}\left(\frac{G}{\pi v_F}\right)^2\left(\frac{M}{m}-1\right),\,\,\,G\ll v_F.
\end{eqnarray}
At $P=k_F$ we instead use $E(P=k_F)\approx\frac{k_F^2}{2M}$ and neglect terms of order $(M-mN)/M^*(P=k_F)\ll1$. Recalling that $M^*(P=k_F)\propto -mG/v_F$, these approximations are seen to be valid for small deviations away from the integrable point, $|1-M/m|<G/v_F\ll1$. Substituting them into Eq.~(\ref{gamma12_Pn}) gives
\begin{eqnarray}
\label{eq:Gamma-weak2}
\Gamma(P=k_F)=-\frac{2\pi}{mv_F^2}\left(\frac{M}{m}-1\right),\,\,\,G\ll v_F.
\end{eqnarray}
In all cases the backscattering amplitude scales as $1-M/m$, thus vanishing at the exactly integrable point $M=m$. This implies the vanishing of the thermal viscosity and the divergence of the Kubo mobility $\sigma_{\mathrm{Kubo}}$ at finite temperature. In this case the response to even an infinitesimal external force is nonlinear: the velocity exhibits Bloch oscillations superimposed with the drift  $V_{\mathrm{D}}=\sigma F$.

\section{Impurity in a trapped condensate}
\label{sec:harmonic}

We now  consider the dynamics of an impurity coupled to a 1D quantum liquid confined by a weak harmonic potential $V(x)=\frac{1}{2}m\omega^2 x^2$ with $\omega\ll\mu$, where $\mu$ is the chemical potential of the 1D gas in the trap center. In this case the spatial extent $L$ of the gas is much larger than the healing length $\xi$ and one may use the local density approximation (LDA), for which the local chemical potential is given by $\mu(x)=\mu-\frac{1}{2}m\omega^2x^2$ for $|x|<\sqrt{\frac{2\mu}{m\omega^2}}\equiv L$, while $\mu(x)=n(x)=0$ for $|x|>L$. In the LDA, one first solves the homogeneous problem at fixed density for the depleton dispersion law $E(P,n)$ and then substitutes in it the local density $n(X)$ to obtain the adiabatic depleton Hamiltonian
\begin{figure}
\includegraphics[width=.48\columnwidth]{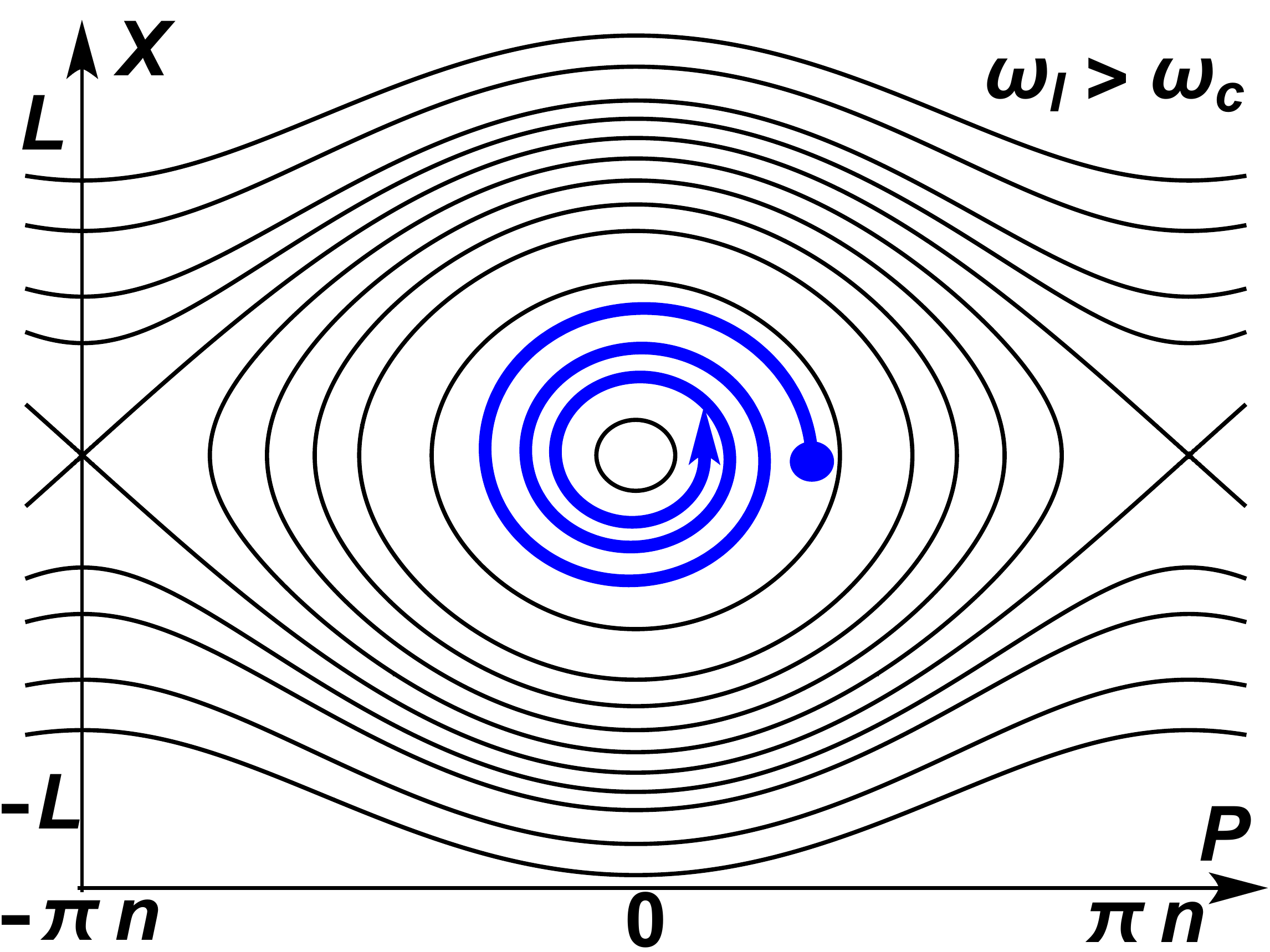}
\includegraphics[width=.48\columnwidth]{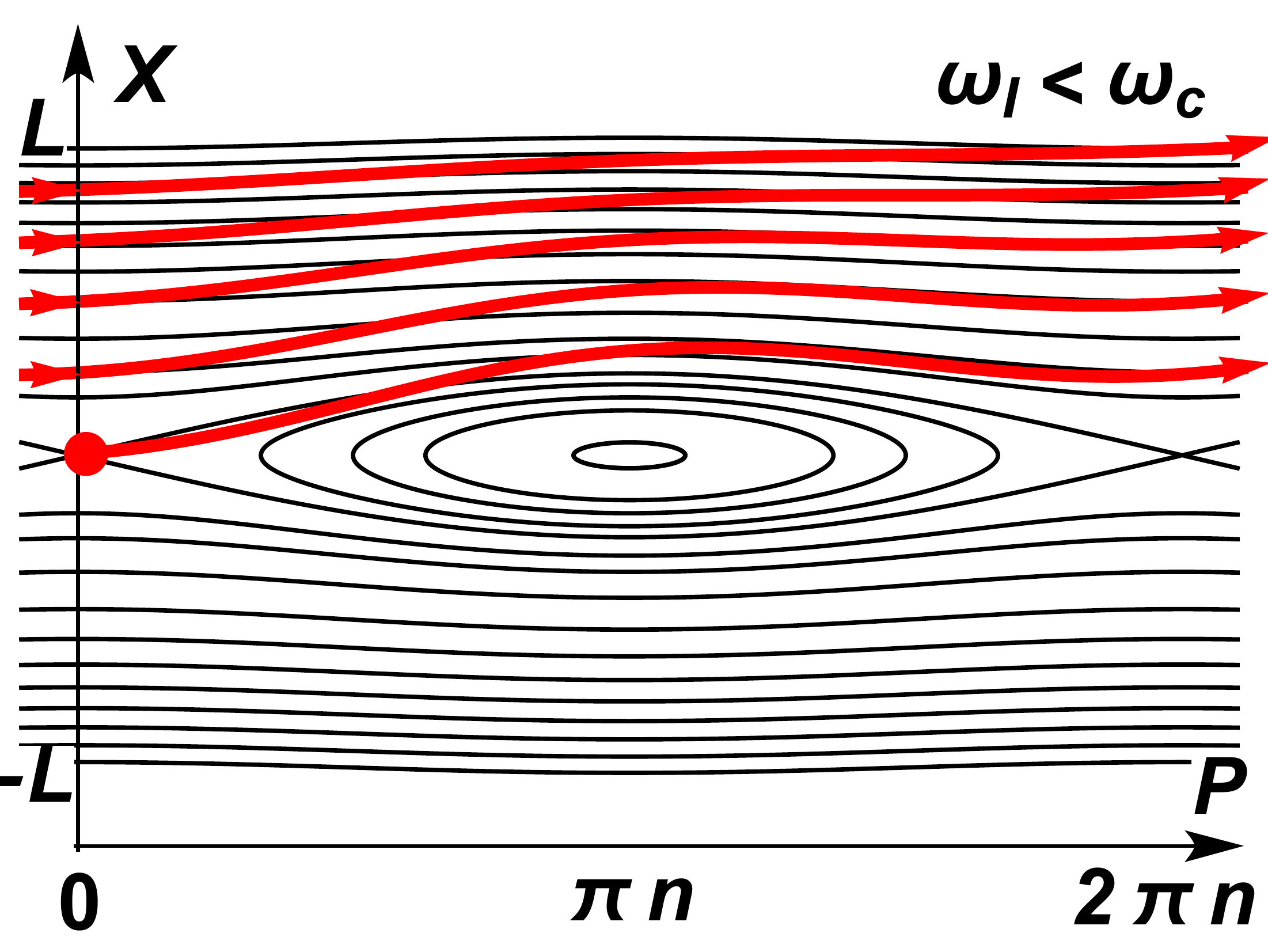}
\caption{(Color online) \Rev{Impurity trajectory in a trapped quantum liquid. The black curves represent schematic constant energy contours of Eq.~(\ref{eq:Hd}). Left panel: for $\omega_I>\omega_c$ the adiabatic orbits are stable in the vicinity of the energy minimum at $(X,P)=(0,0)$. Impurity acceleration leads to phonon damping and thus a decreasing energy and oscillation amplitude (thick blue curve).  Right panel: for $\omega_I<\omega_c$ ($\omega_I=0$ shown) the orbits are instead stable near the energy maximum $(X,P)=(0,\pi n)$. In the case of strong coupling, the maximal displacement on the separatrix orbit is much smaller than the trap size $L$. In this case the impurity escapes by radiating phonons in the running momentum  phase (here $P$ is plotted modulo $2\pi n$), where the velocity exhibits Bloch oscillations plus drift (thick red curve).}}
\label{fig:trap}
\end{figure}
\begin{align}
\label{eq:Hd}
H_{\mathrm{trap}}(P,X)=E(P,n(X))+\frac{1}{2}M\omega_I^2X^2.
\end{align}
Here we introduced an additional harmonic potential, acting on the impurity only, as a control field that can tune the system into different regimes of stability (this can be achieved using e.g. a species or state selective potential).

In the limit $\omega_I\to\infty$ the impurity is strongly localized in the trap center, while for $\omega_I\to0$ it is instead expelled from the center by the repulsive potential produced by the inhomogeneous density profile of the host particles. The transition between these two regimes occurs at a critical value of the trapping frequency, which can be deduced by expanding Eq.~(\ref{eq:Hd}) in small deviations away from $X=P=0$
\begin{align}
\label{eq:Hd1}
H_{\mathrm{trap}}(P,X)\approx \frac{P^2}{2M^*}+\frac{1}{2}\left(M\omega_I^2-mN\omega^2\right)X^2+E_0,
\end{align}
where we used Eq.~(\ref{eq:Phi N}) and defined $E_0=E(0,N)$ (note that the inverse effective mass $1/M^*=\partial^2_P E$ is distinct from $1/\mathcal{M}$ used above, see e.g. Eq.~(\ref{eq:vel1})). From Eq.~(\ref{eq:Hd1}) we see that both the impurity mass and trapping potential are renormalized by interactions with the background particles, and act to make the motion of the impurity slower (generally $N>0$ and $M^*>M$ near $P=0$). The effective oscillation frequency of the impurity is
\begin{align}
\label{eq:osc}
\Omega=\sqrt{\frac{M\omega_I^2-mN\omega^2}{M^*}}<\omega_I.
\end{align}
As one lowers $\omega_I$, the oscillation frequency $\Omega$ decreases and crosses zero at the critical value of the trapping frequency
\begin{align}
\label{eq:osc-crit}
\omega_{c}=\sqrt{\frac{mN}{M}}\,\omega.
\end{align}
 This signals the frequency at which the trap center near $P=0$ becomes an unstable maximum and here $\Omega$ becomes purely imaginary (illustrated by the saddle point in Fig.~\ref{fig:trap} for $\omega_I<\omega_c$). Setting $\omega_I=0$ for simplicity, we see that in the limit of weak coupling, $G\ll c$, the lifetime of the impurity initiated at rest in the trap center can be estimated as $\rm{Im}\,\Omega^{-1}$ using Eq.~(\ref{eq:osc}). This is appropriate because the maximal displacement from the center on the separatrix orbit is  already roughly the trap size: $X\sim\sqrt{\frac{E_\pi-E_0}{mN\omega^2}}\sim L$, ($E_\pi-E_0\sim \mu N$ at weak coupling), thus allowing the impurity to reach the trap edge and escape. 

At strong coupling, however, the maximal displacement on the separatrix is much smaller, $X\sim L\sqrt{c/G}\ll L$, owing to the small impurity bandwidth, $E_\pi-E_0\sim \mu N(c/G)$, see Fig.~\ref{fig:trap}. This implies that the impurity becomes `self-trapped' in a high energy metastable state by the background gas, and can only escape by releasing energy into phonon excitations. This dissipation allows the impurity to rapidly cross the separatrix and enter the running momentum phase, accompanied by a drift towards the trap edge and small amplitude Bloch oscillations in the velocity, shown by the thick red curve in Fig.~\ref{fig:trap}. Here, Bloch oscillations are driven by the gradient of the inhomogeneous density profile of the gas. The timescale and trajectory of the escape can be estimated by noting that since the force is an increasing function of the displacement, $F=-\partial_X H_\mathrm{trap}\sim mN\omega^2X$, the displacement, in turn, satisfies the differential equation $\dot X=\sigma F\sim \frac{\omega^2}{\mu}X$ (at strong coupling $\sigma=\frac{1}{2nmc}$, and one can neglect the amplitude of velocity oscillations). This leads to the exponential increase of the impurity displacement, $X(t)\propto e^{\omega^2 t/\mu}$, on the timescale $\mu/\omega^2\sim 170$ ms for the parameters used in the experiment of Ref.~\cite{Catani2012}, discussed below. We note that in the extreme limit $G\to\infty$ ($V_c\to0$) the impurity cannot escape, since the number of particles in the left and right condensates become conserved quantities. This implies that the lifetime of the trapped impurity sharply increases beyond $\mu/\omega^2$ as a function of coupling, once $mV_c<1/L$. For the system studied in Ref.~\cite{Catani2012} this yields a crossover coupling ($G/g\sim 100$) that greatly exceeds the experimental values, so we do not pursue this special limit further.

The above results can be tested experimentally by localizing an impurity in the center of a trapped gas, and measuring the width of the subsequent impurity distribution $\langle X^2(t)\rangle$ as a function of time. This was done in Ref.~\cite{Catani2012} using a species selective dipole potential to initialize a $^{41}$K impurity in a gas of moderately interacting $^{87}$Rb atoms ($mg\sim n$). The ratio of the trapping frequencies was fixed at $\omega_I/\omega=1.3$, while the K-Rb scattering length was varied by a magnetic field using a Feshbach resonance. From Eq.~(\ref{eq:osc-crit}) we find a critical coupling strength given by $N_c\sim G_c/g=M\omega_I^2/m\omega^2\sim 1$, above which we have $\omega_I<\omega_c$ and below $\omega_I>\omega_c$.

  At stronger coupling, $G>g$, we thus expect the self-trapping behavior to become pronounced, which appears consistent with the results of Ref.~\cite{Catani2012} showing a rapid decrease of the initial oscillation amplitude for $G>g$ (see Fig. 4 of \cite{Catani2012}). The characteristic timescale for the increase of the width at the largest coupling in Ref.~\cite{Catani2012} ($G/g=30$) is a factor of $\sim8$ faster than $\mu/\omega^2$. Aside from a possible numerical prefactor (that goes beyond the accuracy of the above discussion), this discrepancy could also be explained by the fact that the temperature in Ref.~\cite{Catani2012} is rather large $T\sim \mu$, making the thermal dissipation channel highly relevant (the system is far from integrability due to the K-Rb mass difference), thus giving a faster decay time. The high temperature makes a quantitative comparison with Ref.\cite{Catani2012} difficult  since at weak coupling $G\ll g$, $T$ is already substantially larger than the K-Rb interaction energy $nG\ll T$, while at strong coupling $G/c\gg 1$ the temperature is comparable to or exceeds the impurity bandwidth $T>nV_c$. Accessing lower temperatures, or perhaps closeness to integrability (using e.g. internal hyperfine states of Rb) would make a direct quantitative comparison to the above theoretical results possible (see also Ref.~\cite{cug}).

\section{Conclusions and open questions}
\label{sec:concl}

In this paper we have provided an overview of the physics of mobile impurities
in 1D quantum liquids using the simplified mobile Josephson Junction model and
generalizing it to  the phenomenological \emph{depleton} model.  Our
description is based on the existence of the equilibrium dispersion relation
$E(P,n)$, defined as the ground state of the combined system of an impurity
and the superfluid background, at a given momentum $P$ and background
density $n$. This dispersion curve can be understood in terms of the
thermodynamics of a quantum liquid flowing past an impurity. We have exploited
the periodicity properties of the dispersion to predict the existence of
adiabatic Bloch oscillations in the absence of an underlying lattice. The
interaction of the mobile impurity with low energy phononic excitations was
described in terms of two slow collective variables, which allowed us to
address, in particular, the effects of dissipation and dephasing. Based on
these results, we were able to show that the dynamics of impurities in uniform
and trapped systems can be fully characterized.

Using our exact general results, we have provided model-specific calculations
of the linear (Kubo) and non-linear mobilities in the tractable limits of a
weakly interacting and a strongly interacting background. It is
interesting to see that both these limits lead to a clear physical picture of
 a depleton consisting of 
the repulsive impurity binding to an effective ``hole'' in the background. In
this way  the depleton  properties, such as the effective mass, become strongly
interaction and momentum dependent.

A spinless particle interacting with a scalar background represents the
simplest case of a mobile impurity. Including internal degrees of freedom of the
impurity and those of the background particles are expected to change
\emph{qualitatively} the low energy physics, like in the case of spin 1/2
impurity moving in the background made of spin 1/2 fermions 
\cite{PhysRevLett.101.225301}. In this case the spin-spin interactions become
singular at low energy due to the formation of a Kondo polaron and lead to the
mobility behaving as $T^{-2}$ at low temperatures. Extending these studies to
bosonic backgrounds and other values of spin may result in interesting effects
of entanglement and strong correlations which can be probed experimentally by
radio-frequency pulses.
 
Our description was limited to small applied forces and low temperatures, where
the concept of remaining close to the equilibrium zero-temperature dispersion
remains meaningful.  One open question is to understand to what extent our
results apply to the cases of stronger forces or higher temperatures that are
typical of current experiments in ultracold atoms. Another open
question  is the physics of depleton formation relevant at initial 
stages of dynamical experiments with impurities.

\section*{Acknowledgments}
The authors would like to thank A. Lamacraft, M. Zvonarev, M. Knap, E. Demler,
O. Lychkovskiy, O. Gamayun, V. Cheianov, T. Giamarchi, I.V. Lerner and 
 B. Horovits for
many enlightening conversations that have contributed to our understanding of
impurity dynamics in reduced dimensions.  A.K. was supported by NSF grant
DMR1306734. M.S. was supported by the Danish National Research Foundation and
The Danish Council for Independent Research | Natural Sciences. M.S. and
D.G. gratefully acknowledge the hospitality of the University of Minnesota.

\bibliography{library}

\end{document}